\documentclass[reprint,aps,prx,10pt,amsmath,amssymb,amsfonts,floatfix]{revtex4-2}
\usepackage[utf8]{inputenc}
\usepackage{graphicx}
\usepackage{bbm}
\usepackage{mathtools}
\usepackage{xr-hyper}
\usepackage{hyperref}
\hypersetup{colorlinks,citecolor=black,filecolor=black,linkcolor=blue,urlcolor=blue}
\begin{document}

\title{Spatiotemporal noise stabilizes unbounded diversity in strongly-competitive communities}

\author{Amer Al-Hiyasat}
\thanks{These authors contributed equally to this work.}
\author{Daniel W. Swartz}
\thanks{These authors contributed equally to this work.}
\author{Jeff Gore}
\author{Mehran Kardar}
\email{kardar@mit.edu}
\affiliation{Department of Physics, Massachusetts Institute of Technology, Cambridge, Massachusetts 02139, USA}


\begin{abstract}
Classical ecological models predict that diverse communities should be unstable, presenting a central challenge to explaining the stable biodiversity seen in nature. We revisit this long-standing problem by extending the generalized Lotka-Volterra model to include both spatial structure and environmental fluctuations across space and time. We find that neither space nor environmental noise alone can resolve the tension between diversity and stability, but that together they permit arbitrarily many species to stably coexist in a sufficiently large system, despite strongly disordered competitive interactions. We analytically characterize the noise-induced transition to coexistence, showing that spatiotemporal noise drives power-law abundance fluctuations, leading to an anomalous scaling of moments known empirically as Taylor's law. At the metacommunity level, this manifests as an emergent sublinear self-inhibition that stabilizes diversity and renders the interaction disorder irrelevant in the high-diversity limit. Spatiotemporal noise thus provides a novel resolution to the diversity-stability paradox and a generic mechanism by which complex communities can persist.
\end{abstract}

\date{\today}

\maketitle

\begin{figure*}[!t]
    \centering
    \includegraphics{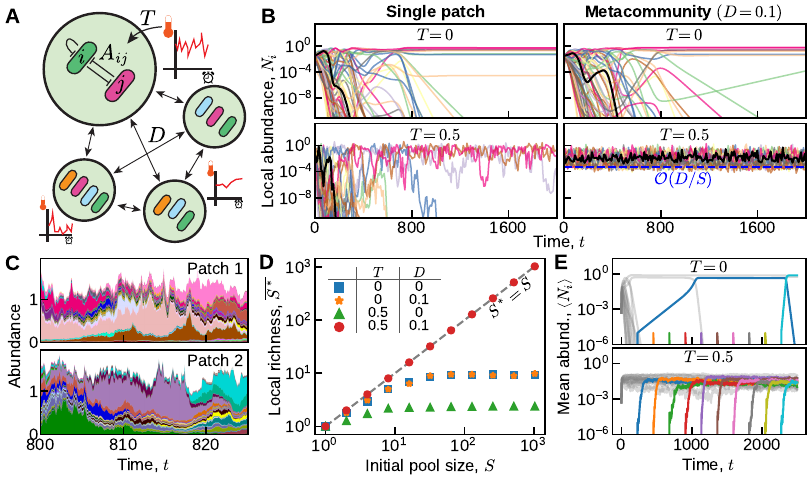}
    \caption{\textbf{Spatiotemporal noise stabilizes unbounded diversity.} (A) Generalized Lotka-Volterra metacommunity with fluctuating growth rates. (B) Abundance trajectories from a single patch system (left column), or a representative patch in a metacommunity (right column), with and without noise (upper and lower rows, respectively). The interaction matrix is identical in all cases and a particular focal species is emphasized in black. Initial pool size is $S = 64$. (C) Stacked species abundances in two representative patches for \(T=0.5, D=0.1\); colors correspond to species. $S=64$. (D) Surviving richness $S^\star$ as a function of initial pool size $S$ for the different combinations of $T$ and $D$ in (B). (E) A metacommunity with $S=16$ is prepared (grey) and new species are introduced at fixed intervals (colored curves), with and without noise. In all panels, $r=1$, $P=2^{14}$, $A_{ij} \sim \mathcal{N}(0.7, 0.2^2)$ and $N_c = 10^{-15}$.}
    \label{fig:fig1}
\end{figure*}

Natural ecosystems are extraordinarily diverse, with hundreds to thousands of species coexisting across scales, from tropical rainforests and coral reefs to microbial communities \cite{wilson1999diversity, knowlton2010coral, curtis2002estimating, qin2010human}. Classical ecological models, however, predict that such diversity is unstable to competitive exclusion~\cite{hardin1960competitive}. May’s diversity–stability paradox sharpens this tension~\cite{may1972will}: in large, randomly-interacting communities, adding species makes stable coexistence increasingly unlikely. Consistent with this prediction, when species that co-occur in nature are placed in well-mixed, controlled laboratory conditions, they often fail to coexist and instead competitively exclude~\cite{gause1934struggle, tilman1977resource, dal2021resource, hu2022emergent}. How, then, is stable biodiversity maintained in nature?

    Several routes around this paradox have been proposed \cite{chesson2000mechanisms, mccann2000diversity}. One approach is to impart a specific structure onto the matrix of interspecific interactions; for example, through sparsity, modularity, or strong correlations \cite{thebault2010stability, allesina2012stability, mougi2012diversity,  rohr2014structural, bairey2016high, grilli2016modularity, grilli2017higher, posfai2017metabolic, pearce2020stabilization, rohr2025will}. Spatial structure provides further mechanisms: if the interaction coefficients vary as much throughout space as they do between species, then different spatial patches select different winners, and diversity can be maintained at the community level~\cite{macarthur1964competition,  levin1974dispersion,chesson1985coexistence, johnson2023coexistence, seidelmann2025species,roy2020complex,garcia2024interactions}. A second approach has been to move beyond static theories, where stationary abundances correspond to stable equilibria of a dynamical system, and instead take account of temporal fluctuations~\cite{connell1978diversity, chesson1985coexistence, roy2020complex, pearce2020stabilization, yachi1999biodiversity, ferraro2026will}. Environmental variability, due to fluctuations in abiotic factors such as weather or nutrient availability, has been suggested to create temporal niches that favor coexistence through a ``storage effect" if it acts asymmetrically on different species~\cite{chesson1981environmental, chesson1994multispecies, holt1996chaotic, chesson2000mechanisms, Lai2005noise, d2008biodiversity, picoche2019self, pande2022temporal, burkart2023periodic, johnson2023coexistence, van2024tiny, mallmin2025fluctuating, asker2025fixation, seidelmann2025species}. However, in the disordered competitive setting most directly tied to May’s argument, environmental noise alone does not generically stabilize coexistence~\cite{turelli1978does, fox2013intermediate} and can instead accelerate diversity loss by driving rare species toward extinction~\cite{lande1993risks}.

    A standard setting in which May's argument is borne out is the generalized Lotka–Volterra (gLV) model with random interactions~\cite{bunin2017ecological}, which reproduces key dynamical features of laboratory microcosms~\cite{hu2022emergent, amor2022fast, hu2025collective}.
    In the recent theoretical literature, it has been typical to focus on weakly-interacting communities, where interaction coefficients are made to vanish as the inverse of the species pool size~\cite{kokkoris1999patterns, bunin2017ecological, biroli2018marginally, altieri2021properties, garcia2024interactions}. Although this scaling can generate diverse stable equilibria, it implies that, in rich communities, pairwise cross-inhibition is negligible compared to self-inhibition, effectively endowing each species with its own distinct niche. This is at odds with direct measurements in microbial systems~\cite{hu2022emergent}, and is least plausible precisely where the diversity–stability paradox is most acute—for instance, the coexistence of thousands of very similar phytoplankton on only a handful of resources~\cite{hutchinson1961paradox}.

    In this work, we demonstrate that two ubiquitous features of natural ecosystems, (1) spatial structure and (2) spatiotemporal environmental noise, suffice to stabilize unbounded coexistence in large, strongly-competitive metacommunities. Working in the randomly-interacting generalized Lotka-Volterra framework, we show that neither ingredient alone stabilizes diversity, but that their combination creates a new phase in which the number of coexisting species is limited only by the size of the metacommunity.
    The mechanism of this noise-induced transition leaves clear macroecological footprints with empirical backing: spatiotemporal noise generates heavy-tailed abundance distributions~\cite{ottino2020population, swartz2022seascape, mallmin2025fluctuating}, giving rise to a manifestation of Taylor's law~\cite{taylor1961aggregation, eisler2008fluctuation}, where abundance fluctuations scale as an anomalous power of the mean. This, in turn, leads to an emergent nonanalytic, sublinear self-inhibition at the community level, which we show to stabilize unbounded diversity~\cite{hatton2024diversity, mazzarisi2024complexity}. {Within the coexistence phase, the effects of interaction heterogeneity diminish as diversity is increased, so that the dynamics of diverse communities are largely insensitive to the details of interactions. } These macroecological patterns are not imposed as modeling choices; they emerge together with the stabilization of richness, strengthening the plausibility of our framework.

\section{\NoCaseChange{g}LV metacommunities with spatiotemporal noise}
    To model spatiotemporal noise, we augment the gLV model as follows (Fig.~\ref{fig:fig1}a): we implement spatial structure using a metacommunity, comprising a network of ``patches" with local interactions within each patch and dispersal between them~\cite{hanski1998metapopulation, leibold2004metacommunity, pearce2020stabilization, garcia2024interactions, denk2022self}. We include environmental noise through fluctuations in the species growth rates, which amounts to a multiplicative noise term proportional to the species abundance. This is to be distinguished from demographic noise, which scales as the square root of the abundance. The abundance of species $i \in 1, \dots, S$ on patch $\alpha \in 1, \dots, P$ then follows
\begin{align}
    \label{eq:abundance_full}
\dot{N}_{i\alpha} =&
    r N_{i\alpha} \bigg[ 1 - N_{i\alpha} - \!\sum_{j\neq i} A_{ij}N_{j\alpha} \bigg] \nonumber \\
    &\!+ \frac{D}{c} \! \sum_{\beta \in \partial \alpha} \!(N_{i\beta} - N_{i\alpha}) + \sqrt{2T} N_{i\alpha} \eta_{i \alpha}(t),
\end{align}
where $\partial \alpha$ denotes the neighbors of patch $\alpha$, and $c$ is the number of neighbors, assumed uniform across patches. The parameters $D$ and $T$ set the dispersal rate and strength of environmental noise, respectively. For simplicity and in line with prior literature~\cite{garcia2024interactions, de2025self, mallmin2025fluctuating}, the same $D$ and $T$ are used for all species.
 We take $\eta_{i\alpha}(t)$ to be a unit Gaussian white noise\footnote{Interpreted in the It\^o sense.}, independent among species and patches,
\begin{equation}
    \langle \eta_{i\alpha}(t) \eta_{j\beta}(t') \rangle = \delta_{ij} \delta_{\alpha \beta} \delta(t-t').
\end{equation}
In the above, $\langle \cdot \rangle$ denotes an ``ensemble" average over noise realizations, $\delta_{ij}$ and $\delta_{\alpha \beta}$ denote Kronecker deltas, and $\delta(t-t')$ is the Dirac delta. The interaction matrix, $A$, is uniform across patches and constant in time (quenched).
Its elements are independent and identically distributed with finite mean and variance:
\begin{equation} \label{eq:Astatistics} \overline{A_{ij}} = \mu,\qquad \overline{(A_{ij} A_{k\ell})}_c = \delta_{ik}\delta_{j \ell} \sigma^2.
\end{equation}
Here, $\overline{(\cdot)}$ denotes an average over the realizations of $A$, equivalent to a species average for large pool size, $S$. Since $\mu$ and  $\sigma^2$ are $\mathcal{O}(1)$ rather than $\mathcal{O}(1/S)$, the ecological communities considered here are strongly interacting~\cite{garcia2022well, mallmin2024chaotic}. 
To model demographic extinction, we impose a cutoff $N_c \ll 1/S$ below which $N_{i\alpha}(t)$ is set to zero, which is valid for large but finite populations in which demographic stochasticity is insignificant~\cite{ovaskainen2010stochastic}.

To make analytical progress, we focus on fully-connected networks of size $P \gg S \gg 1$, though we also consider the extension to finite networks and spatial lattices. The dynamics then take the form~\cite{ottino2020population, swartz2022seascape}
\begin{align}
    \label{eq:abundance_mf}
\dot{N}_{i} =&
    r N_{i} \bigg[1\! - \! N_{i} \!- \!\sum_{j\neq i}\! A_{ij}N_{j} \bigg]\!+\! D (\left\langle N_{i}\right \rangle \!-\! N_{i}) \!+\! \sqrt{2T} N_{i} \eta_{i},
\end{align}
where the patch index has been dropped and $\langle N_i \rangle$ denotes the patch-averaged abundance $\langle N_i \rangle \equiv \frac{1}{P}\sum_\alpha N_{i\alpha}$, equivalent to an ensemble average in the limit $P\rightarrow \infty$.

\section{Noise-induced transition to coexistence}
We performed numerical simulations of Eq.~\eqref{eq:abundance_mf} in the presence or absence of spatial structure or environmental noise (Fig.~\ref{fig:fig1}B). In absence of both (Fig.~\ref{fig:fig1}B, top left), most species abundances flow towards the extinction cutoff, with only a handful of survivors at long times. Adding space or noise alone does not alter this behavior (Fig.~\ref{fig:fig1}B, top right and bottom left). Surprisingly, however, when the dispersal rate is nonzero and the noise is sufficiently strong, all species persist indefinitely, with abundances fluctuating between a migration floor set by $D \langle N_i \rangle/r$, and an $\mathcal{O}(1)$ ceiling set by the carrying capacity (Fig.~\ref{fig:fig1}B, bottom right).  Notably, we find that the patch-averaged abundances, $\langle N_i \rangle$, are $\mathcal{O}(1/S)$. Spatiotemporal noise thus stabilizes a coexistence state in which an $\mathcal{O}(1)$ total biomass is shared among $S$ species (Fig.~\ref{fig:fig1}C). {Though every species is extant everywhere, the bulk of local biomass is carried by a small, fluctuating subset of temporarily dominant species, with the remaining species sitting near the $\mathcal{O}(1/S)$ floor. This dominant subset is different across patches and turns over dynamically in time (Fig.~\ref{fig:fig1}C)~\cite{henderson2026fluctuating, mallmin2024chaotic, mallmin2025fluctuating}. Diversity is thus sustained through a dynamic ``mass" effect~\cite{shmida1985biological,leibold2004metacommunity}, in which dispersal maintains species locally while the identities of source and sink patches change over time.}

To quantify how much diversity can be stabilized by spatiotemporal noise, we measured the local richness, $S^\ast$, defined as the average number of long-time survivors on one patch (for $D>0$, local richness equals global richness, as any extant species is present on all patches). This is plotted in Fig~\ref{fig:fig1}D as a function of the initial pool size, $S$, for a large metacommunity with $P \gg S$. In the standard single-patch case ($T=D=0$), consistent with known results \cite{bunin2017ecological, may1972will}, the surviving richness initially increases with $S$ but eventually saturates, implying an asymptotic survival fraction of $\phi \equiv \lim_{S\rightarrow\infty} S^\ast/S = 0$. Environmental noise ($T>0, D=0$) lowers the upper bound on $S^\ast$, showing that environmental noise alone in fact favors extinction\footnote{Environmental fluctuations alone can stabilize coexistence in models with overlapping generations, where long-lived individuals carry populations through unfavorable periods~\cite{chesson1981environmental}. The gLV has no such mechanism~\cite{turelli1978does, fox2013intermediate, lande1993risks}.}. Without noise but with nonzero dispersal ($T=0, D>0$), the limiting value of $S^\ast$ remains $\mathcal{O}(1)$~\cite{roy2020complex}, so that the survival fraction is still zero for large $S$ (unless $A_{ij}$ is sufficiently antisymmetric, as reported in Ref.~\cite{pearce2020stabilization}. We do not consider that choice here as it is not required for coexistence). With both space and noise, however, the surviving richness grows indefinitely with $S$ as $S^\ast = S$, implying coexistence of every species from the initial pool and a survival fraction of unity. Crucially, the interaction parameters $\mu$ and $\sigma^2$ do not vanish with increasing $S$,  so that richer communities are not stabilized by weakening interactions.

Unbounded coexistence is insensitive to the value of the extinction cutoff, so long as the cutoff is well-separated from the migration floor $N_c \ll 1/S$ [Supplementary information (SI), Fig.~\ref{fig:cutoff}]. Coexistence is also robust to temporal correlations or small spatial correlations in environmental noise (SI, Figs.~\ref{fig:temporalNoise} and~\ref{fig:spatialNoise}). For finite $P$,
the number of survivors, $S^\ast(S, P)$,  initially grows with $S$ before saturating at some maximal $S^\ast_\mathrm{max}(P)$.
For the parameters used in Fig.~\ref{fig:fig1}, $S^\ast_\mathrm{max}$ is of the order $10^2$ when $P = 16$ (SI, Fig.~\ref{fig:finitePScaling}a). The maximal sustainable richness grows without bound as the metacommunity size is increased, though the scaling of $S^\ast_\mathrm{max}(P)$ with $P$ appears nonuniversal (SI, Fig.~\ref{fig:finitePScaling}b). Spatiotemporal noise thus stabilizes a degree of richness which is limited only by the number of independently fluctuating spatial patches.

An illustrative consequence of this result is provided in Fig.~\ref{fig:fig1}E: Here, new species are periodically introduced at random points in space and made to interact with the resident community. In absence of noise (top panel), most attempted invasions fail, and those that succeed typically displace a resident species. When spatiotemporal noise is sufficiently strong, however, all invasions succeed and only minimally disturb the resident community.

\begin{figure}[!t]
    \centering
    \includegraphics[width=\columnwidth]{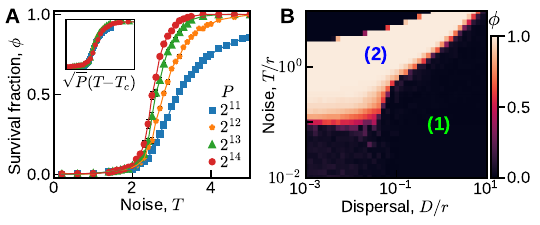}
    \caption{\textbf{Noise-induced transition to coexistence.} (A) Survival fraction $\phi = S^\star/S$ as a function of noise strength $T$, showing a transition from $\phi=0$ to $\phi=1$ which sharpens as $P$ is increased. Inset shows collapse of the curves near $T_c$ upon rescaling with $\sqrt{P}$. $D=r=1, S = 512$.
    (B) Survival fraction $\phi$ as a function of dispersal rate and noise magnitude, showing different phases of community diversity. $S=256, P=2^{14}$. In both panels, $A_{ij} \sim \mathcal{N}(2, 1), N_c = 10^{-15}$.}
    \label{fig:critical_behavior}
\end{figure}

We next asked how strong the noise must be to stabilize this high-diversity state. We find that there is a critical noise strength, $T_c$, below which the survival fraction approaches zero and above which it approaches unity. The transition becomes sharp in the infinite $P$ limit, with a width that vanishes as $1/\sqrt{P}$ (Fig.~\ref{fig:critical_behavior}A, inset). To determine how $T_c$ depends on the dispersal rate, $D$, we plot in Fig.~\ref{fig:critical_behavior}B the survival fraction as a function of both $T$ and $D$. We identify three phases: when the noise is weak (phase 1), the metacommunity is in a low diversity competitive exclusion phase with an $\mathcal{O}(1)$ number of survivors, so that $\phi \rightarrow 0$ for large $S$. As $T$ is increased, the system transitions to a noise-stabilized coexistence phase with a survival fraction of unity (phase 2). If $T$ is increased further still, so that it is significantly larger than $r$, the diversity collapses again and even a single species cannot survive due to the large noise ($S^\ast = 0$). Since no species survive in this phase, it is not relevant to the question of competitive exclusion, and we hereafter focus only on phases (1) and (2) and the transition line between them $T_c(D)$. We also note that the large-$T$ extinction phase vanishes if $\log P \gg T-r$ (SI), and is thus absent from sufficiently large metacommunities.

The coexistence phase spans a broad range of dispersal rates: when $D \ll r$, the critical noise strength, $T_c(D)$, approaches a constant independent of $D$. The limit $D\rightarrow 0$ is thus singular, since there is no transition in absence of dispersal. In this $D \ll r$ regime, where few migration events happen within a generation time, $T_c$ is of the order $10^{-1}r$, implying that growth rate fluctuations of as little as $\sim 10\%$ suffice to stabilize coexistence. Increasing the dispersal rate delays the coexistence transition: for $D \gg r$, the critical noise strength scales linearly as $T_c(D) \propto D$. As we soon show, this is because the patches evolve synchronously when $D$ is large, behaving effectively as a single-patch gLV system, unless the noise is strong enough to desynchronize them (as in Fig.~\ref{fig:fig1}C). Coexistence is thus most readily stabilized at nonzero but small dispersal rates.

In the following, we develop an analytical theory, formally valid in the opposite limit $D \gg r$, to uncover the mechanism behind the noise-induced coexistence phase. Although derived in this high-migration regime, our theory captures several qualitative features of the coexistence phase across a broad range of dispersal rates, down to $D \ll r$, as verified in the SI (section~\ref{sec:smallD}).

\begin{figure*}[t]
    \centering
    \includegraphics{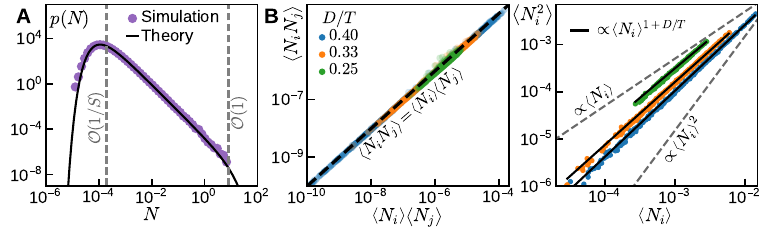}
    \caption{\textbf{Power law abundance distributions and emergence of Taylor's law}. (a) Truncated power law distribution over patches for a particular species, showing agreement with Eq.~\eqref{eq:powerlaw}. $S=256$. (b) Verification of the closure $\langle N_{i} N_j \rangle = \langle N_i \rangle \langle N_j \rangle$ (left) and the anomalous scaling $\langle N_i^2 \rangle \propto \langle N_i \rangle^{1+D/T}$ (right), demonstrating the emergence of Taylor's law. Data combines simulations with $S = 128$ and $S=256$ for the three indicated values of $D/T$. In all panels, $P = 2^{22}$, $A_{ij} \sim \mathcal{N}(4, 4)$, and $D = 2r$.}
    \label{fig:richards}
\end{figure*}

\section{Effective dynamics of mean abundances}
Averaging Eq.~\eqref{eq:abundance_mf} yields the following dynamics for the mean (patch-averaged) abundances:
\begin{equation} \label{eq:meanDynamics}
    \dot{\langle N_i\rangle} = r\bigg[\langle N_i \rangle - \langle N_i^2 \rangle - \sum_{j\neq i} A_{i j} \langle N_i N_j \rangle \bigg],
\end{equation}
where we have used the limit $P\rightarrow \infty$ to equate patch and ensemble averages. Since the dynamics of the means depend on the correlators $\langle N_i N_j \rangle$, Eq.~\eqref{eq:meanDynamics} is not closed in general.
As we show in the \textit{Methods}, however, a closure may be obtained in the limit $D \gg r$: The correlator $\langle N_i N_{j\neq i} \rangle$ becomes a fast variable relaxing on a time scale $D^{-1}$ and can be adiabatically eliminated in favor of the means $\{ \langle N_i \rangle \}$, which evolve slowly on a time scale $r^{-1}$ (Eq.~\ref{eq:meanDynamics}). This yields
\begin{equation} \label{eq:offDiagClosure}
    \langle N_i N_{j\neq i} \rangle = \langle N_i\rangle \langle N_j \rangle + \mathcal{O}\left(\textstyle{\frac{r}{D}}\right).
\end{equation}
Whereas Eq.~\eqref{eq:offDiagClosure} holds irrespective of the value of $T$, the closure for $\left\langle N_i^2 \right\rangle$ instead exhibits a transition at $T = D$ that portends the onset of the coexistence phase. For $T<D$, we find
\begin{equation} \label{eq:glvclosure}
\langle N_i^2 \rangle = \frac{D}{D-T}\langle N_i \rangle^2 + \mathcal{O}\big(\textstyle{\frac{r}{D-T}}\big), \qquad (T<D).
\end{equation}
Substituting into Eq.~\eqref{eq:meanDynamics} then yields an effective (single-patch) gLV dynamics for the means, reducing the effect of spatiotemporal noise to a renormalization of the carrying capacity. This places the region $T<D$ within the low-diversity competitive exclusion phase.

To understand the coexistence phase, we require a closure for $\langle N_i^2 \rangle$ which is valid for $T>T_c \geq D$. To achieve this, we study Eq.~\eqref{eq:abundance_mf} under the conditions that: (1) all $S$ species coexist with $\langle N_i \rangle > 0$, (2) the mean abundances are of order $1/S$, and (3) the means are held adiabatically fixed in time because $D\gg r$. Conditions (1) and (2) will be verified as self-consistent above a critical noise strength that will be determined shortly and identified with $T_c$. {Subject to these conditions, our first result, derived in the \textit{Methods}, is that the interaction term $\sum_j A_{ij} N_j$ has negligible fluctuations over time or across species, becoming sharply peaked around unity in the large-$S$ limit.
This is reminiscent of self-organized criticality~\cite{denk2022self, de2025self, calvo2026microbiome}, in that the interactions drive the effective linear growth rate to zero without fine-tuning.
The dynamics then reduce to an effective single-species problem, whose stationary solution is the exponentially truncated power law (Fig.~\ref{fig:richards}a):}

\begin{equation} \label{eq:powerlaw} p(N_i) \propto e^{-rN_i/T} e^{-D \langle N_i \rangle/(T N_i)} N_i^{-2-D/T},
\end{equation}
 The lower cutoff at $N_i \sim D \langle N_i \rangle/T \propto 1/S$ constitutes the migration floor observed in Fig.~\ref{fig:fig1} and becomes arbitrarily small as $S\rightarrow \infty$. Equation~\eqref{eq:powerlaw} can be interpreted as a distribution of abundances across patches at a given time, or as a distribution over time on a single patch. Power law abundance distributions have been reported broadly in the empirical literature \cite{locey2016scaling, ser2018ubiquitous, grilli2020macroecological, gao2025powerbend} with a range of exponents. We note that the form in Eq.~\eqref{eq:powerlaw} was previously derived by some of us in the single-species setting~\cite{ottino2020population, swartz2022seascape}, and, very recently, reported in neutral multi-species models~\cite{mallmin2025fluctuating}. That it survives in the strongly-disordered competitive setting studied here is a novel result. The most important consequence of Eq.~\eqref{eq:powerlaw} is that it leads to an anomalous scaling of moments \cite{ottino2020population, swartz2022seascape}:
\begin{equation} \label{eq:taylor} \langle N_i^2 \rangle \propto \left \langle N_i \right \rangle^{1+ D/T}, \qquad (T>T_c).
\end{equation}
This constitutes an instance of \textit{Taylor's law}, the widely verified empirical observation that the variance of species abundance scales as a power of the mean abundance \cite{taylor1961aggregation, eisler2008fluctuation}. Spatiotemporal noise thus provides a possible mechanistic explanation of this empirical law, with an exponent determined by the noise strength and dispersal rate. We note that Eq.~\eqref{eq:taylor} in fact generalizes to all higher moments as $\langle N^{1+p} \rangle \propto \langle N \rangle ^{1+ D/T}$ \cite{ottino2020population, swartz2022seascape}. This implies that if the logistic self-regulation term of the gLV model, $-N^2$, is replaced with an arbitrary analytic term $-N f(N)$, spatiotemporal noise renormalizes this to a universal anomalous form at the community level: $\langle N f(N) \rangle \propto \langle N \rangle^{1+ D/T}$.

We verify Eq.~\eqref{eq:taylor}, together with Eq.~\eqref{eq:offDiagClosure}, in large-scale simulations reported in Fig.~\ref{fig:richards}B. Although our analytical results should in principle apply only when $D\gg r$, numerical data shows excellent agreement even at $D = 2r$. When $D\ll r$, Eq.~\eqref{eq:offDiagClosure} breaks down as expected. Taylor's law, however, remains robust, but with an exponent different from $1+D/T$ and which cannot be determined from the present large-$D$ theory (Fig.~\ref{fig:smallD}).

Combining Eq.~\eqref{eq:taylor} and Eq.~\eqref{eq:offDiagClosure}, we obtain the following effective dynamics for the mean abundances:
\begin{equation} \label{eq:effectiveDynamics}
    \langle \dot{N_i} \rangle = r \langle N_i \rangle \bigg[1- K^{-\theta} \langle N_i \rangle^{\theta} - \sum_{j\neq i} A_{ij} \langle N_j \rangle \bigg],
\end{equation}
where the effective carrying capacity $K(D,T)$ is given in the SI. The Taylor's law exponent $\theta$ is equal to unity for $T<D$, whereas in the coexistence phase $T>T_c$ we have $\theta = D/T$. The growth law in Eq.~\eqref{eq:effectiveDynamics} is often referred to as $\theta$-logistic~\cite{sibly2005regulation, hatton2024diversity} or Richard's \cite{richards1959flexible, swartz2022seascape}
growth, and we thus hereafter refer to the interacting model as a $\theta$-logistic gLV ($\theta$-gLV) equation. Previous attempts to fit empirical data have introduced this model phenomenologically \cite{gilpin1973global, hatton2024diversity}; here, we show that it emerges as a natural and universal consequence of spatiotemporal noise.

\begin{figure*}
    \centering
    \includegraphics[width=\textwidth]{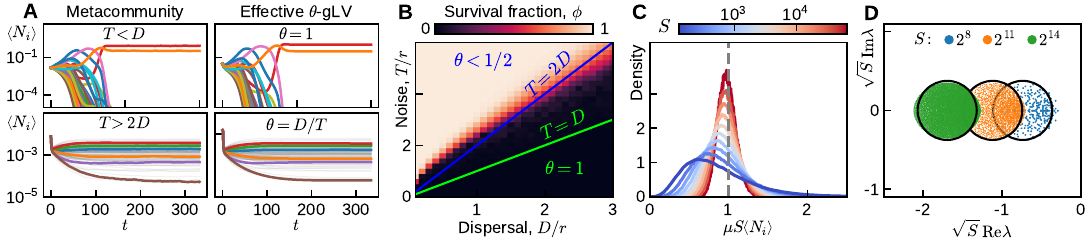}
    \caption{\textbf{Effective $\theta$-logistic gLV dynamics of mean abundances}. (a) Patch-averaged abundance trajectories from metacommunity simulations (left) compared to the prediction of Eq.~\eqref{eq:effectiveDynamics} (right). For clarity, six species are emphasized within a background of $256$ in the coexistence case $T>2D$. In the top panels, $D = 8r$ and $T = 2r$, whereas in the bottom $D = 2r$ and $T = 6r$. (b) Survival fraction as a function of $(T, D)$ obtained from metacommunity simulations and plotted on a linear axis. The boundaries at $T=D$ and $T=2D$ are indicated, representing the breakdown of effective gLV dynamics followed by the transition to coexistence ($\theta<1/2$). (c) Distribution (over species) of patch-averaged abundances, obtained from simulations of the effective $\theta$-gLV dynamics. The distribution narrows as $S$ is increased. (d) Stability spectrum of the $\theta$-gLV model: each point is an eigenvalue of the fixed-point Jacobian, obtained numerically from one simulation at each indicated pool size, $S$. Black circles show the prediction of Eq.~\eqref{eq:stability}. Parameter values are given in the SI.}
    \label{fig:thetamapping}
\end{figure*}

Equation~(\ref{eq:effectiveDynamics}) reduces the stochastic metacommunity with $S\times P$ variables into a single-patch deterministic system of only $S$ variables, making predictions that are valid at the single-trajectory level, as verified in Fig.~\ref{fig:thetamapping}a.
To understand how the $\theta$-gLV model allows for stable coexistence, we now analyze its fixed points and study their stability, from which we determine the critical noise threshold for the coexistence transition, $T_c$.

\section{Fixed points of the $\theta$-\NoCaseChange{g}LV model}
In the steady state, the mean abundances will obey Eq.~\eqref{eq:effectiveDynamics} with  $\langle \dot{N}_i \rangle = 0$. It can then be shown using standard techniques \cite{hatton2024diversity} that a typical $\langle N_i \rangle$ will, for large $S$, converge in distribution to the random variable
\begin{equation} \label{eq:cavityDist} \langle N \rangle = K \max{\left(0, 1-\mu S \overline{\langle N \rangle} -  \sigma \sqrt{S \overline{\langle N \rangle^2}} \,Z\right)}^{1/\theta},
\end{equation}
where $Z \sim \mathcal{N}(0,1)$ is a standard Gaussian variable representing the randomness in $A$. The species averages $\overline{(\cdot)}$ are then obtained by averaging over $Z$, and the survival fraction is $\phi \equiv \mathbb{P}[\langle N \rangle >0]$. We show in the \textit{Methods} that the self-consistent system can be solved exactly for large $S$, revealing a transition to a novel coexistence phase when $\theta$ crosses $1/2$ from above:
\begin{equation} \label{eq:criticaltheta}
    \phi = \begin{cases}
        0, &\theta>1/2\\
        1, &\theta<1/2.
    \end{cases}
\end{equation}
Notably, for $\mu, \sigma > 0$, the coexistence condition has no dependence on $\mu$ or $\sigma$; this is qualitatively different from the weakly-interacting gLV model, where coexistence is possible only below a maximal degree of disorder, $\sigma$. In the special case of uniform interactions ($\sigma=0$) with $\mu>1$, the coexistence criterion is weakened to $\theta<1$ (SI). Together with the $D \gg r$ solution $\theta = D/T$, Eq.~\eqref{eq:criticaltheta} predicts the critical noise strength for coexistence in the metacommunity to be $T_c = 2D$ for any $\mu, \sigma > 0$, a result consistent with our numerical phase diagram (Fig.~\ref{fig:thetamapping}b). This also proves that the region $D<T<2D$ cannot support a coexistence phase, though the detailed dynamics in this region cannot be inferred from Eq.~\eqref{eq:effectiveDynamics}, whose derivation for $T>D$ relies on coexistence.

Within the coexistence phase, the distribution (over species) of patch-averaged abundances is characterized by the moments $\overline{\langle N \rangle}$ and $\overline{\langle N \rangle ^2}$, which we obtain perturbatively for large $S$ as
\begin{equation} \label{eq:summary} \overline{\langle N \rangle} \simeq \frac{1}{\mu S}\left[1 - \frac{1}{(K \mu S)^\theta} \right], \qquad \overline{\langle N \rangle^2} \simeq \frac{1}{\mu^2 S^2}.
 \end{equation}
This explains our numerical finding that $\langle N_i \rangle \propto 1/S$. Furthermore, since $\overline{\langle N \rangle^2} \rightarrow \overline{\langle N\rangle}^2$ as $S \rightarrow \infty$, it predicts that all patch-averaged abundances $\langle N_i \rangle$ will, for large $S$, become sharply peaked around their species average:
\begin{equation} \label{eq:coexistenceFP} \langle N_i \rangle \xrightarrow{S\gg1} \overline{\langle N \rangle} = \frac{1}{\mu S}.
\end{equation}
{This result, confirmed in Fig.~\ref{fig:thetamapping}C, implies that interaction disorder is asymptotically irrelevant for determining patch-averaged abundances in the coexistence phase. Since these averages are the only quantities that distinguish species in Eq.~\eqref{eq:powerlaw}, we conclude that species become identically distributed for large-$S$, as in a neutral model. This suggests a possible route by which neutral phenomenology can emerge in communities with strongly disordered interactions, offering a connection to the empirical success of neutral models in ecology \cite{hubbell2001unified, bell2001neutral, loreau2008species, kalyuzhny2015neutral, pande2022temporal}.}

The irrelevance of interaction disorder in the large-$S$ limit extends to coexisting metacommunities with $D\ll r$, even though the mapping to the $\theta$-gLV is no longer exact in that regime (SI, Fig.~\ref{fig:smallD}B). That the interactions are strong is essential: with weak interactions $\mu, \sigma^2 \propto 1/S$, mean abundances are $\mathcal{O}(1)$ rather than $\mathcal{O}(1/S)$, and simulations then show that the distribution of $\langle N_i \rangle$ has a large-$S$ limit of finite width (SI, Fig.~\ref{fig:weakSAD}B). We furthermore note that sources of interspecific heterogeneity other than interaction disorder can remain relevant in the large-$S$ limit; for example, variations in bare growth rates (SI, Fig.~\ref{fig:randomRSAD}A).

To see how the high diversity fixed point identified in Eq.~\eqref{eq:coexistenceFP} evades the diversity-stability problem, we linearize Eq.~\eqref{eq:effectiveDynamics} about this fixed point. Writing $\langle N_i \rangle \equiv 1/\mu S + \delta\langle N_i \rangle$, we obtain the linearized dynamics
\begin{equation} \label{eq:stability}
    \delta \langle \dot{\mathbf{N}} \rangle = -\frac{r}{\mu S}\left[\theta K^{-\theta} (\mu S)^{1-\theta} \mathbbm{1} +A\right]\delta \langle \mathbf{N} \rangle,
\end{equation}
where $\mathbbm{1}$ denotes the identity. The stability matrix has negative diagonal entries scaling as $S^{-\theta}$ and random off-diagonals $\propto 1/S$. It then follows from standard results that, apart from a negative outlier associated with the uniform mode, the bulk eigenvalues lie in a disk in the complex plane, centered at $-r \theta (\mu S K)^{-\theta}$ and with radius $r \sigma/\mu \sqrt{S}$. For large-$S$, the fixed point is thus almost surely stable if and only if $\theta<1/2$, confirming the stability of coexistence when $T > 2D$. This is illustrated in Fig.~\ref{fig:thetamapping}, which shows the distribution of eigenvalues shifting towards stability when diversity is increased, as reported for the related model in Ref.~\cite{hatton2024diversity}.

\section{Discussion}
    Almost every natural environment is spatially structured and fluctuates throughout space and time. In this work, we have shown that these two ingredients suffice to resolve the tension between diversity and stability in precisely the setting that has confounded the field since May's seminal work: species-rich ecosystems with dense, heterogeneous interactions that do not vanish with $S$. Our mechanism makes no appeal to specific patterns of sparsity, correlation, or (anti)symmetry in the interaction matrix. We do not claim such structure is absent in real systems, only that it is not necessary for stable coexistence.

    Our choice to implement environmental noise through fluctuations in the bare growth rates was motivated by parsimony and prior literature, but is not unique \cite{asker2025fixation}. Indeed, diversity can be enhanced if the interaction matrix, $A$, itself fluctuates throughout space~\cite{garcia2024interactions, roy2020complex} or time~\cite{suweis2024generalized, zanchetta2025emergence}. In these cases, however, to stabilize extensive diversity requires the fluctuations in $A$ to be comparable to the fixed part of $A$. The latter is expected to dominate the former in many settings, since ecological interactions are strongly constrained by intrinsic species traits (e.g., prey do not become predators). In contrast, single-species growth rates often fluctuate widely~\cite{hairston1996overlapping}.

    The mechanism of the transition to coexistence is intimately tied to the macroecological consequences of spatiotemporal fluctuations: noise generates power-law abundance distributions, which imply Taylor's law of fluctuation scaling. At the community level, this manifests as an effective $\theta$-logistic self-inhibition in the dynamics of the mean abundances. The emergent $\theta$-gLV model accounts for the coexistence transition and bears similarities to the sublinear growth model analyzed in Ref.~\cite{hatton2024diversity}, which is itself a $\theta$-gLV with $\theta < 0$. Here, however, the underlying model is entirely analytic in the abundances, and it is spatiotemporal fluctuations that renormalize the dynamics into the nonanalytic $\theta$-gLV equation. Unlike in the $\theta<0$ model, the per-capita growth rate remains finite at low abundance for $\theta>0$, making the stabilization of diversity less intuitive.
    
    \nocite{de2024many}
    Beyond our analytical mapping to the $\theta$-gLV model, a simpler physical picture is that environmental noise produces broad, desynchronized abundance fluctuations across patches (Fig.~\ref{fig:fig1}C), allowing a species that is rare in most places to become transiently abundant somewhere and thus maintaining the migration floor above extinction~\footnote{Power laws have also been reported in noiseless gLV models with externally imposed abundance floors~\cite{mallmin2024chaotic,de2024many}, where they arise intrinsically from chaotic ecological dynamics with $A$ as in Eq.~\eqref{eq:Astatistics}.  Understanding why these fluctuations fail to sustain a self-consistent migration floor in a metacommunity (Figs.~\ref{fig:fig1} and ~\ref{fig:critical_behavior}) while extrinsic environmental fluctuations succeed, remains an open problem; our theory addresses only the large-$D$ regime through Eqs.~(\ref{eq:glvclosure}-\ref{eq:offDiagClosure}).}.
In the language of Modern Coexistence Theory~\cite{chesson2000mechanisms}, spatiotemporal noise generates positive fitness-density covariance, allowing a rare invader with a negative patch-averaged fitness to grow by accumulating in patches where its fitness is transiently higher (SI, Fig.~\ref{fig:gdcov}; see also Ref.~\cite{johnson2023coexistence}).

    In finite metacommunities within the coexistence phase, the surviving richness is limited only by system size: the maximal sustainable richness, $S^\ast_{\max}(P)$, diverges with $P$ (SI, Fig.~\ref{fig:finitePScaling}). Determining how $S^\ast_{\max}(P)$ depends on $P$ and other model parameters lies beyond the infinite-$P$ theory developed here, but would connect our results to empirical species--area relationships and the implications of habitat loss for biodiversity. Even at infinite $P$, not all variants of our model display full coexistence of every species from the initial pool: Preliminary simulations show that quenched heterogeneity in intrinsic growth rates can drive some species extinct despite strong noise (SI, Fig.~\ref{fig:growthRateSurvival}). In weakly-interacting metacommunities, the discontinuous transition from $\phi=0$ to $\phi=1$ is replaced by a continuous crossover from $\phi<1$ to $\phi=1$ as $T$ is increased (SI, Fig.~\ref{fig:weakSurvival}). A counterintuitive consequence is that, in some parameter regimes, increasing the mean competition strength, $\mu$, can \textit{increase} the survival fraction by moving a weakly-interacting system into the strongly-interacting coexistence phase (SI, Fig.~\ref{fig:phiOfmu}).

   An important simplifying assumption made in this work is the fully-connected network topology. To make quantitative comparison with empirical data on terrestrial or marine ecosystems will require the extension of our results to two- or three-dimensional spaces~\cite{swartz2022seascape, munoz1998nonlinear, tu1997systems}. We leave this interesting theoretical problem to future work, but present, in the SI (Fig.~\ref{fig:finitedim}), preliminary numerical evidence for a coexistence transition in two-dimensional lattices. Beyond modifications to the spatial network, our work invites several natural extensions relevant to real ecosystems, such as investigating the role of interspecific correlations in environmental noise or heterogeneity in dispersal rates and noise strengths across species.

    From a broader physics perspective, most work on noisy gLV dynamics has focused on demographic noise, proportional to $\sqrt{N}$, rather than environmental noise, which scales as $N$ \cite{altieri2021properties, denk2022self, garcia2024interactions, de2025self}. The former has the advantage of satisfying a fluctuation-dissipation theorem (FDT), so that by choosing the interaction matrix to be symmetric, the system resembles an equilibrium spin glass and can be treated using standard techniques~\cite{altieri2021properties, garcia2024interactions}. Environmental noise, in contrast, violates the FDT, driving the system out of equilibrium regardless of the symmetry of the interactions. This allows the system to escape the low-diversity glassy phase even with strong interactions, which is impossible with demographic noise alone. Our work thus connects ecological coexistence to the broader literature on nonequilibrium glass transitions~\cite{fisher1988nonequilibrium,berthier2013non}.

    On the empirical side, our framework challenges the standard practice of inferring interaction networks by fitting bare gLV models. Indeed, unobserved spatial fluctuations could masquerade as neutrality in the interaction matrix, confounding estimates derived from well-mixed models. At the same time, our results offer a potential simplification: we identify the emergent self-regulation exponent, $\theta$, as the control parameter for the transition to coexistence. Because $\theta$ manifests macroecologically through Taylor's law, it can be estimated from single-species abundance data, implying that the persistence of a complex ecosystem could be diagnosed without reconstructing the full interaction matrix. If borne out in experimental microcosms, this prediction would establish Taylor’s law as a probe of dynamical stability.

\section{Materials and Methods}
\subsection*{Adiabatic approximation for $D \gg r$}
To obtain a closure for Eq.~\eqref{eq:meanDynamics}, we consider the dynamics of the second moments, which can be derived from Eq.~\eqref{eq:abundance_mf} as
\begin{align}
    \partial_t \langle N_i^2 \rangle =& 2D \left(\langle N_i \rangle^2 - \langle N_i^2 \rangle \right) + 2T \langle N_i^2 \rangle  \label{eq:variance} \\ &+ 2r\bigg[\langle N_i^2\rangle - \langle N_i^3 \rangle -\textstyle{\sum_{j\neq i}} A_{ij} \langle N_i^2 N_j \rangle \bigg], \nonumber\\
    \partial_t \langle N_i N_j \rangle =& 2D\left(\langle N_i \rangle \langle N_j \rangle - \langle N_i N_j \rangle\right) + \mathcal{O}(r), \label{eq:twopoint}
\end{align}
where we have used Ito's lemma. Every term in Eq.~\eqref{eq:meanDynamics} is proportional to $r$, indicating a relaxation time proportional to $1/r$,
whereas Eq.~\eqref{eq:twopoint} relaxes on a time proportional to $1/2D$. In the limit $r \ll D$, we may thus set $\partial_t \langle N_i N_{j \neq i} \rangle = 0$, which yields Eq.~\eqref{eq:offDiagClosure}. A similar closure can be obtained for $\langle N_i^2 \rangle$ only when $T<D$: Eq.~\eqref{eq:variance} relaxes on a time $\sim 1/2(D-T)$, so that $\langle N_i^2 \rangle$ becomes a fast variable when $r \ll D - T$, yielding Eq.~\eqref{eq:glvclosure}. The mean abundances then follow an effective single-patch gLV dynamics for $T < D-r$, with a reduced carrying capacity equal to $K=1-T/D$. Though this is equivalent to a weakening of interactions (as can be seen through the rescaling $N \rightarrow N/K$), it does not suffice to stabilize unbounded diversity.

\subsection*{Power law abundance distribution}
Starting with Eq.~\eqref{eq:abundance_mf}, we define the competitive load on species $i$ as
\begin{equation} \label{eq:xidef} \xi(t) \equiv \sum_{j\neq i} A_{ij} N_j.
\end{equation}
We show here that in the coexistence phase and for $D\gg r$, $\xi_i(t)$ has negligible fluctuations over time or between species. To do so, we replace $\xi_i(t)$ with its stationary mean $\xi_i \simeq \langle \xi_i \rangle$, and later show self-consistently that $\langle \xi_i^2 \rangle_c \simeq 0$. With this replacement, the Fokker-Planck equation corresponding to Eq.~\eqref{eq:abundance_full} reads
\begin{equation} \label{eq:FPE} \partial_t p = -\partial_N \{[N(1 - \langle \xi \rangle-N)+D(\langle N\rangle - N)]p\} + T\partial_N^2 \left[N^2 p\right],\end{equation}
where we have dropped the species index and set $r$ to unity by rescaling time. In the adiabatic limit $T,D \gg 1$, the mean abundance $\langle N \rangle$ is quasistatically fixed, and higher moments relax quickly to values determined by the stationary solution to Eq.~\eqref{eq:FPE}, which takes the form of a power law with exponential cutoffs~\cite{ottino2020population, swartz2022seascape, mallmin2025fluctuating}:
\begin{equation} \label{eq:statsol}
p(N) \propto e^{-\beta N} e^{-\beta D\langle N\rangle/N} N^{-\gamma}, \qquad \gamma \equiv 2 + \beta(D+\langle \xi \rangle-1).
\end{equation}
where $\beta \equiv T^{-1}$ denotes the inverse noise strength. The moments of Eq.~\eqref{eq:statsol} can be expressed exactly in terms of Bessel functions and then characterized asymptotically for small $\langle N_i \rangle \propto 1/S$. This approach is detailed in the SI. Here, we instead rely on scaling arguments to approximate the effects of the cutoffs. We demand self-consistently that $\langle N \rangle = \int d N N p(N)$. For this to hold as $\langle N \rangle \rightarrow 0$ requires $\gamma > 2$, so that the integral is dominated by the lower cutoff and approaches
\[\langle N \rangle \simeq \frac{\int_{0}^\infty dN e^{-\beta D \langle N \rangle/N} N^{1-\gamma}}{\int_{0}^\infty dN e^{-\beta D \langle N \rangle/N} N^{-\gamma}} = \frac{\beta D \langle N \rangle}{\gamma - 2}.\]
Self-consistency then imposes $\gamma - 2 = \beta D$, implying the large-$S$ result
\begin{equation} \label{eq:xilabel}
\langle \xi \rangle = 1 + o(S^0). \end{equation}
Notably, this is the same for all species and exactly cancels the linear growth term in Eq.~\eqref{eq:statsol}, constituting a form of self-organized criticality~\cite{de2025self, denk2022self} and showing that the effect of disordered interactions simplifies significantly in the large-$S$ limit of the coexistence phase. We show in the SI that the subleading corrections to Eq.~\eqref{eq:xilabel} scale as $S^{-\beta D}$ and differentiate the species.

\subsection*{Taylor's law}
The second moment of Eq.~\eqref{eq:statsol} is given by the integral $\langle N \rangle = \int N^2 p(N)$, which is dominated by the lower cutoff for $\gamma > 3$ and the upper cutoff for $\gamma < 3$, where $\gamma = 2 + \beta D$ according to Eq.~\eqref{eq:xilabel}. For the former case, where $T<D$, we have
\[\langle N^2 \rangle \simeq \frac{\int_{0}^\infty dN e^{-\beta D \langle N \rangle/N} N^{-\beta D}}{\int_{0}^\infty dN e^{-\beta D \langle N \rangle/N} N^{-2-\beta D}} = \frac{D}{D - T} \langle N \rangle^2\]
which recovers the adiabatic result. For $T>D$, the correct large-$S$ scaling can be obtained by replacing the exponential cutoffs with a hard lower cutoff proportional to $\langle N \rangle \in \mathcal{O}(S^{-1})$ and an $\mathcal{O}(1)$ upper cutoff. It then follows, for any $p\geq 1$, that when $T>D$
\[\langle N^{1+p} \rangle \sim \frac{\int_{\langle N \rangle}^{\mathcal{O}(1)} dN \, N^{-\beta D}}{\int_{\langle N \rangle}^{\mathcal{O}(1)} dN \, N^{-2-\beta D}} \propto \langle N \rangle^{1+\beta D}\]
The exact prefactor is obtained for $p=1$ in the SI using Bessel function expansions. This establishes the transition to anomalous scaling reported in Eq.~\eqref{eq:taylor}.

The results above, together with the adiabatic factorization approximation of Eq.~\eqref{eq:offDiagClosure}, suffice to check the self-consistency of our replacement of $\xi_i(t)$ with its mean in the coexistence phase: it can be verified that $\langle \xi_i^2 \rangle = \sum_{k, j \neq i} A_{ij} A_{ik} \langle N_i N_j \rangle = \langle \xi \rangle ^2 + \mathcal{O}(S^{-\beta D})$.

\subsection*{Cavity solution of the strongly-interacting $\theta$-\NoCaseChange{g}LV model}
Equation~\eqref{eq:cavityDist} can be written in the form $\langle N \rangle = K \max(0, m+g Z)^{1/\theta}$, where
\[m \equiv 1 - \mu S \overline{\langle N \rangle}, \qquad g^2 \equiv \sigma^2 S \overline{\langle N \rangle^2}.\]
and $Z \sim \mathcal{N}(0,1)$. The moments $\overline{\langle N \rangle}$ and $\overline{\langle N \rangle^2}$ then obey the self-consistency conditions
\begin{equation} \label{eq:cavitySelfCon} \overline{\langle N \rangle^n} = K^{n} \int_{-m/g}^\infty \! \! Dz\, (m + g z)^{n/\theta},\end{equation}
where $Dz \equiv dz\, e^{-z^2/2}/\sqrt{2\pi}$. The survival fraction is given by $\phi = \int_{-m/g}^\infty Dz$. Searching for a coexistence phase, we demand that $\lim_{S\rightarrow \infty} m/g = \infty$. We then invoke the large-$S$ scaling ansatz
\begin{equation} \label{eq:scalingAnsatz} \overline{\langle N \rangle} \sim S^{-\alpha}, \quad \overline{\langle N^2 \rangle} \sim S^{-\gamma}, \quad \frac{m}{g} \sim S^{\eta},\end{equation}
for some $\alpha, \gamma, \eta > 0$. Substituting into Eq.~\eqref{eq:cavitySelfCon} and expanding for large $S$ then allows the self-consistent equations to be solved perturbatively. This is implemented to leading order in the SI, yielding exponent identities which may be solved to find $\alpha = 1, \gamma = 2$ and $\eta = \frac{1}{2} - \theta$. Since coexistence requires $\eta > 0$, we obtain the criterion in Eq.~\eqref{eq:criticaltheta}. The $\mathcal{O}(1)$ prefactors in Eq.~\eqref{eq:scalingAnsatz} can then be computed as detailed in the SI, yielding Eq.~\eqref{eq:summary}.

\section*{Acknowledgments}
We are grateful to Yizhou Liu for an early derivation of Eq.~\eqref{eq:criticaltheta} in the weakly-interacting case. We thank Thibaut Arnoulx de Pirey, Jiliang Hu, and Julien Tailleur for helpful discussions. A.A. acknowledges the support of the Tushar Shah and Sarah Zion Physics Fellowship, as well as a grant from the Institute for Complex Adaptive Matter. M.K. was supported by the NSF through Grant No. DMR-2218849.

\bibliography{bibliography.bib}

@article{hu2022emergent,
  title={Emergent phases of ecological diversity and dynamics mapped in microcosms},
  author={Hu, Jiliang and Amor, Daniel R and Barbier, Matthieu and Bunin, Guy and Gore, Jeff},
  journal={Science},
  volume={378},
  number={6615},
  pages={85--89},
  year={2022},
  publisher={American Association for the Advancement of Science}
}

@article{bunin2017ecological,
  title={Ecological communities with {Lotka-Volterra} dynamics},
  author={Bunin, Guy},
  journal={Physical Review E},
  volume={95},
  number={4},
  pages={042414},
  year={2017},
  publisher={APS}
}

@article{altieri2021properties,
  title={Properties of equilibria and glassy phases of the random {Lotka-Volterra} model with demographic noise},
  author={Altieri, Ada and Roy, Felix and Cammarota, Chiara and Biroli, Giulio},
  journal={Physical Review Letters},
  volume={126},
  number={25},
  pages={258301},
  year={2021},
  publisher={APS}
}

@article{taylor1961aggregation,
  title={Aggregation, variance and the mean},
  author={Taylor, Lionel Roy},
  journal={Nature},
  volume={189},
  number={4766},
  pages={732--735},
  year={1961},
  publisher={Springer Nature}
}

@article{eisler2008fluctuation,
  title={Fluctuation scaling in complex systems: Taylor's law and beyond},
  author={Eisler, Zolt{\'a}n and Bartos, Imre and Kert{\'e}sz, J{\'a}nos},
  journal={Advances in Physics},
  volume={57},
  number={1},
  pages={89--142},
  year={2008},
  publisher={Taylor \& Francis}
}

@book{hubbell2001unified,
  title={The Unified Neutral Theory of Biodiversity and Biogeography (MPB-32)},
  author={Hubbell, Stephen P},
  volume={32},
  year={2001},
  publisher={Princeton University Press}
}

@article{bell2001neutral,
  title={Neutral macroecology},
  author={Bell, Graham},
  journal={Science},
  volume={293},
  number={5539},
  pages={2413--2418},
  year={2001},
  publisher={American Association for the Advancement of Science}
}

@article{garcia2024interactions,
  title={Interactions and migration rescuing ecological diversity},
  author={Garcia Lorenzana, Giulia and Altieri, Ada and Biroli, Giulio},
  journal={PRX Life},
  volume={2},
  number={1},
  pages={013014},
  year={2024},
  publisher={APS}
}

@article{roy2020complex,
  title={Complex interactions can create persistent fluctuations in high-diversity ecosystems},
  author={Roy, Felix and Barbier, Matthieu and Biroli, Giulio and Bunin, Guy},
  journal={PLoS computational biology},
  volume={16},
  number={5},
  pages={e1007827},
  year={2020},
  publisher={Public Library of Science San Francisco, CA USA}
}

@article{pearce2020stabilization,
  title={Stabilization of extensive fine-scale diversity by ecologically driven spatiotemporal chaos},
  author={Pearce, Michael T and Agarwala, Atish and Fisher, Daniel S},
  journal={Proceedings of the National Academy of Sciences},
  volume={117},
  number={25},
  pages={14572--14583},
  year={2020},
  publisher={National Academy of Sciences}
}

@article{swartz2022seascape,
  title={Seascape origin of Richards growth},
  author={Swartz, Daniel W and Ottino-L{\"o}ffler, Bertrand and Kardar, Mehran},
  journal={Physical Review E},
  volume={105},
  number={1},
  pages={014417},
  year={2022},
  publisher={APS}
}

@article{ottino2020population,
  title={Population extinction on a random fitness seascape},
  author={Ottino-L{\"o}ffler, Bertrand and Kardar, Mehran},
  journal={Physical Review E},
  volume={102},
  number={5},
  pages={052106},
  year={2020},
  publisher={APS}
}

@article{mccann2000diversity,
  title={The diversity--stability debate},
  author={McCann, Kevin Shear},
  journal={Nature},
  volume={405},
  number={6783},
  pages={228--233},
  year={2000},
  publisher={Nature Publishing Group UK London}
}

@article{amor2022fast,
  title={Fast growth can counteract antibiotic susceptibility in shaping microbial community resilience to antibiotics},
  author={Amor, Daniel R and Gore, Jeff},
  journal={Proceedings of the National Academy of Sciences},
  volume={119},
  number={15},
  pages={e2116954119},
  year={2022},
  publisher={National Academy of Sciences}
}

@article{may1972will,
  title={Will a large complex system be stable?},
  author={May, Robert M},
  journal={Nature},
  volume={238},
  number={5364},
  pages={413--414},
  year={1972},
  publisher={Nature Publishing Group UK London}
}

@article{rohr2014structural,
  title={On the structural stability of mutualistic systems},
  author={Rohr, Rudolf P and Saavedra, Serguei and Bascompte, Jordi},
  journal={Science},
  volume={345},
  number={6195},
  pages={1253497},
  year={2014},
  publisher={American Association for the Advancement of Science}
}

@article{hatton2024diversity,
  title={Diversity begets stability: Sublinear growth and competitive coexistence across ecosystems},
  author={Hatton, Ian A and Mazzarisi, Onofrio and Altieri, Ada and Smerlak, Matteo},
  journal={Science},
  volume={383},
  number={6688},
  pages={eadg8488},
  year={2024},
  publisher={American Association for the Advancement of Science}
}

@article{yachi1999biodiversity,
  title={Biodiversity and ecosystem productivity in a fluctuating environment: the insurance hypothesis},
  author={Yachi, Shigeo and Loreau, Michel},
  journal={Proceedings of the National Academy of Sciences},
  volume={96},
  number={4},
  pages={1463--1468},
  year={1999},
  publisher={The National Academy of Sciences}
}

@article{hu2025collective,
  title={Collective dynamical regimes predict invasion success and impacts in microbial communities},
  author={Hu, Jiliang and Barbier, Matthieu and Bunin, Guy and Gore, Jeff},
  journal={Nature Ecology \& Evolution},
  volume={9},
  number={3},
  pages={406--416},
  year={2025},
  publisher={Nature Publishing Group UK London}
}

@article{chesson2000mechanisms,
  title={Mechanisms of maintenance of species diversity},
  author={Chesson, Peter},
  journal={Annual review of Ecology and Systematics},
  volume={31},
  number={1},
  pages={343--366},
  year={2000},
  publisher={Annual Reviews 4139 El Camino Way, PO Box 10139, Palo Alto, CA 94303-0139, USA}
}

@article{biroli2018marginally,
  title={Marginally stable equilibria in critical ecosystems},
  author={Biroli, Giulio and Bunin, Guy and Cammarota, Chiara},
  journal={New Journal of Physics},
  volume={20},
  number={8},
  pages={083051},
  year={2018},
  publisher={IOP Publishing}
}

@article{kokkoris1999patterns,
  title={Patterns of species interaction strength in assembled theoretical competition communities},
  author={Kokkoris, GD and Troumbis, AY and Lawton, JH},
  journal={Ecology Letters},
  volume={2},
  number={2},
  pages={70--74},
  year={1999},
  publisher={Wiley Online Library}
}

@article{hardin1960competitive,
  title={The competitive exclusion principle: an idea that took a century to be born has implications in ecology, economics, and genetics.},
  author={Hardin, Garrett},
  journal={Science},
  volume={131},
  number={3409},
  pages={1292--1297},
  year={1960},
  publisher={American Association for the Advancement of Science}
}

@article{hutchinson1961paradox,
  title={The paradox of the plankton},
  author={Hutchinson, G Evelyn},
  journal={The American Naturalist},
  volume={95},
  number={882},
  pages={137--145},
  year={1961},
  publisher={Science Press}
}

@article{turelli1978does,
  title={Does environmental variability limit niche overlap?},
  author={Turelli, Michael},
  journal={Proceedings of the National Academy of Sciences},
  volume={75},
  number={10},
  pages={5085--5089},
  year={1978}
}

@article{lande1993risks,
  title={Risks of population extinction from demographic and environmental stochasticity and random catastrophes},
  author={Lande, Russell},
  journal={The American Naturalist},
  volume={142},
  number={6},
  pages={911--927},
  year={1993},
  publisher={University of Chicago Press}
}

@article{mallmin2025fluctuating,
  title={Fluctuating growth rates link turnover and unevenness in species-rich communities},
  author={Mallmin, Emil and Traulsen, Arne and De Monte, Silvia},
  journal={arXiv preprint arXiv:2505.01376},
  year={2025}
}

@article{mallmin2024chaotic,
  title={Chaotic turnover of rare and abundant species in a strongly interacting model community},
  author={Mallmin, Emil and Traulsen, Arne and De Monte, Silvia},
  journal={Proceedings of the National Academy of Sciences},
  volume={121},
  number={11},
  pages={e2312822121},
  year={2024},
  publisher={National Academy of Sciences}
}

@article{sibly2005regulation,
  title={On the regulation of populations of mammals, birds, fish, and insects},
  author={Sibly, Richard M and Barker, Daniel and Denham, Michael C and Hone, Jim and Pagel, Mark},
  journal={Science},
  volume={309},
  number={5734},
  pages={607--610},
  year={2005},
  publisher={American Association for the Advancement of Science}
}

@article{richards1959flexible,
  title={A flexible growth function for empirical use},
  author={Richards, Francis J},
  journal={Journal of experimental Botany},
  volume={10},
  number={2},
  pages={290--301},
  year={1959},
  publisher={Oxford University Press}
}

@article{gilpin1973global,
  title={Global models of growth and competition},
  author={Gilpin, Michael E and Ayala, Francisco J},
  journal={Proceedings of the National Academy of Sciences},
  volume={70},
  number={12},
  pages={3590--3593},
  year={1973}
}

@article{grilli2020macroecological,
  title={Macroecological laws describe variation and diversity in microbial communities},
  author={Grilli, Jacopo},
  journal={Nature communications},
  volume={11},
  number={1},
  pages={4743},
  year={2020},
  publisher={Nature Publishing Group UK London}
}

@article{gao2025powerbend,
  title={The powerbend distribution provides a unified model for the species abundance distribution across animals, plants and microbes},
  author={Gao, Yingnan and Abdullah, Ahmed and Wu, Martin},
  journal={Nature Communications},
  volume={16},
  number={1},
  pages={4035},
  year={2025},
  publisher={Nature Publishing Group UK London}
}

@article{suweis2024generalized,
  title={Generalized lotka-volterra systems with time correlated stochastic interactions},
  author={Suweis, Samir and Ferraro, Francesco and Grilletta, Christian and Azaele, Sandro and Maritan, Amos},
  journal={Physical Review Letters},
  volume={133},
  number={16},
  pages={167101},
  year={2024},
  publisher={APS}
}

@article{berthier2013non,
  title={Non-equilibrium glass transitions in driven and active matter},
  author={Berthier, Ludovic and Kurchan, Jorge},
  journal={Nature Physics},
  volume={9},
  number={5},
  pages={310--314},
  year={2013},
  publisher={Nature Publishing Group UK London}
}

@article{fisher1988nonequilibrium,
  title={Nonequilibrium dynamics of spin glasses},
  author={Fisher, Daniel S and Huse, David A},
  journal={Physical Review B},
  volume={38},
  number={1},
  pages={373},
  year={1988},
  publisher={APS}
}

@article{dal2021resource,
  title={Resource--diversity relationships in bacterial communities reflect the network structure of microbial metabolism},
  author={Dal Bello, Martina and Lee, Hyunseok and Goyal, Akshit and Gore, Jeff},
  journal={Nature ecology \& evolution},
  volume={5},
  number={10},
  pages={1424--1434},
  year={2021},
  publisher={Nature Publishing Group UK London}
}

@book{gause1934struggle,
  title={The struggle for existence},
  author={Gause, George Francis},
  year={1934},
  publisher={Williams \& Wilkins Baltimore}
}

@article{tilman1977resource,
  title={Resource competition between plankton algae: an experimental and theoretical approach},
  author={Tilman, David},
  journal={Ecology},
  volume={58},
  number={2},
  pages={338--348},
  year={1977},
  publisher={Wiley Online Library}
}

@article{van2024tiny,
  title={A tiny fraction of all species forms most of Nature: Rarity as a sticky state},
  author={van Nes, Egbert H and Pujoni, Diego GF and Shetty, Sudarshan A and Straatsma, Gerben and de Vos, Willem M and Scheffer, Marten},
  journal={Proceedings of the National Academy of Sciences},
  volume={121},
  number={2},
  pages={e2221791120},
  year={2024},
  publisher={National Academy of Sciences}
}

@article{fox2013intermediate,
  title={The intermediate disturbance hypothesis should be abandoned},
  author={Fox, Jeremy W},
  journal={Trends in ecology \& evolution},
  volume={28},
  number={2},
  pages={86--92},
  year={2013},
  publisher={Elsevier}
}

@article{rohr2025will,
  title={Will a large complex model ecosystem be viable? The essential role of positive interactions},
  author={Rohr, Rudolf P and Bersier, Louis-f{\'e}lix and Arditi, Roger},
  journal={Ecology},
  volume={106},
  number={3},
  pages={e70064},
  year={2025},
  publisher={Wiley Online Library}
}

@article{asker2025fixation,
  title={Fixation and extinction in time-fluctuating spatially structured metapopulations},
  author={Asker, Matthew and Swailem, Mohamed and T{\"a}uber, Uwe C and Mobilia, Mauro},
  journal={arXiv preprint arXiv:2504.08433},
  year={2025}
}

@article{denk2022self,
  title={Self-consistent dispersal puts tight constraints on the spatiotemporal organization of species-rich metacommunities},
  author={Denk, Jonas and Hallatschek, Oskar},
  journal={Proceedings of the National Academy of Sciences},
  volume={119},
  number={26},
  pages={e2200390119},
  year={2022},
  publisher={National Academy of Sciences}
}

@article{posfai2017metabolic,
  title={Metabolic trade-offs promote diversity in a model ecosystem},
  author={Posfai, Anna and Taillefumier, Thibaud and Wingreen, Ned S},
  journal={Physical review letters},
  volume={118},
  number={2},
  pages={028103},
  year={2017},
  publisher={APS}
}

@article{zanchetta2025emergence,
  title={Emergence of ecological structure and species rarity from fluctuating metabolic strategies},
  author={Zanchetta, Davide and Gupta, Deepak and Moschin, Sofia and Suweis, Samir and Maritan, Amos and Azaele, Sandro},
  journal={PRX Life},
  volume={3},
  number={3},
  pages={033016},
  year={2025},
  publisher={APS}
}

@book{gardiner2004handbook,
  title={Handbook of stochastic methods},
  author={Gardiner, Crispin W and others},
  volume={3},
  year={2004},
  publisher={springer Berlin}
}

@article{de2025self,
  title={Self-organized criticality in complex model ecosystems},
  author={de Pirey, Thibaut Arnoulx},
  journal={arXiv preprint arXiv:2512.06961},
  year={2025}
}

@misc{bendavid2025submitphysicsanalysisfacility,
    title={{SubMIT: A Physics Analysis Facility at MIT}},
    author={Josh Bendavid and Mariarosaria D'Alfonso and Jan Eysermans and Chad Freer and Maxim Goncharov and Matthew Heine and Luca Lavezzo and Marianne Moore and Christoph Paus and Xuejian Shen and David Walter and Zhangqier Wang},
    year={2025},
    eprint={2506.01958},
    archivePrefix={arXiv},
    primaryClass={cs.DC},
    url={https://arxiv.org/abs/2506.01958},
}

@article{besard2018juliagpu,
  author        = {Besard, Tim and Foket, Christophe and De Sutter, Bjorn},
  title         = {Effective Extensible Programming: Unleashing {Julia} on {GPUs}},
  journal       = {IEEE Transactions on Parallel and Distributed Systems},
  year          = {2018},
  doi           = {10.1109/TPDS.2018.2872064},
  ISSN          = {1045-9219},
  archivePrefix = {arXiv},
  eprint        = {1712.03112},
  primaryClass  = {cs.PL},
}

@book{wilson1999diversity,
  title={The diversity of life},
  author={Wilson, Edward O},
  year={1999},
  publisher={WW Norton \& Company}
}

@article{knowlton2010coral,
  title={Coral reef biodiversity},
  author={Knowlton, Nancy and Brainard, Russell E and Fisher, Rebecca and Moews, Megan and Plaisance, Laetitia and Caley, M Julian and others},
  journal={Life in the world's oceans: diversity distribution and abundance},
  pages={65--74},
  year={2010},
  publisher={Wiley-Blackwell Hoboken, NJ}
}

@article{curtis2002estimating,
  title={Estimating prokaryotic diversity and its limits},
  author={Curtis, Thomas P and Sloan, William T and Scannell, Jack W},
  journal={Proceedings of the National Academy of Sciences},
  volume={99},
  number={16},
  pages={10494--10499},
  year={2002},
  publisher={National Academy of Sciences}
}

@article{qin2010human,
  title={A human gut microbial gene catalogue established by metagenomic sequencing},
  author={Qin, Junjie and Li, Ruiqiang and Raes, Jeroen and Arumugam, Manimozhiyan and Burgdorf, Kristoffer Solvsten and Manichanh, Chaysavanh and Nielsen, Trine and Pons, Nicolas and Levenez, Florence and Yamada, Takuji and others},
  journal={Nature},
  volume={464},
  number={7285},
  pages={59--65},
  year={2010},
  publisher={Nature Publishing Group UK London}
}

@article{grilli2016modularity,
  title={Modularity and stability in ecological communities},
  author={Grilli, Jacopo and Rogers, Tim and Allesina, Stefano},
  journal={Nature communications},
  volume={7},
  number={1},
  pages={12031},
  year={2016},
  publisher={Nature Publishing Group UK London}
}

@article{allesina2012stability,
  title={Stability criteria for complex ecosystems},
  author={Allesina, Stefano and Tang, Si},
  journal={Nature},
  volume={483},
  number={7388},
  pages={205--208},
  year={2012},
  publisher={Nature Publishing Group UK London}
}

@article{grilli2017higher,
  title={Higher-order interactions stabilize dynamics in competitive network models},
  author={Grilli, Jacopo and Barab{\'a}s, Gy{\"o}rgy and Michalska-Smith, Matthew J and Allesina, Stefano},
  journal={Nature},
  volume={548},
  number={7666},
  pages={210--213},
  year={2017},
  publisher={Nature Publishing Group UK London}
}

@article{ovaskainen2010stochastic,
  title={Stochastic models of population extinction},
  author={Ovaskainen, Otso and Meerson, Baruch},
  journal={Trends in ecology \& evolution},
  volume={25},
  number={11},
  pages={643--652},
  year={2010},
  publisher={Elsevier}
}

@article{rosenzweig1995species,
  title={Species diversity in space and time},
  author={Rosenzweig, Michael L},
  journal={(No Title)},
  year={1995},
  publisher={Cambridge university press}
}

@article{Lai2005noise,
  title = {Noise Promotes Species Diversity in Nature},
  author = {Lai, Ying-Cheng and Liu, Yi-Rong},
  journal = {Phys. Rev. Lett.},
  volume = {94},
  issue = {3},
  pages = {038102},
  numpages = {4},
  year = {2005},
  month = {Jan},
  publisher = {American Physical Society},
  doi = {10.1103/PhysRevLett.94.038102},
  url = {https://link.aps.org/doi/10.1103/PhysRevLett.94.038102}
}

@article{hanski1998metapopulation,
  title={Metapopulation dynamics},
  author={Hanski, Ilkka},
  journal={Nature},
  volume={396},
  number={6706},
  pages={41--49},
  year={1998},
  publisher={Nature Publishing Group UK London}
}

@article{holt1996chaotic,
  title={Chaotic population dynamics favors the evolution of dispersal},
  author={Holt, Robert D and McPeek, Mark A},
  journal={The American Naturalist},
  volume={148},
  number={4},
  pages={709--718},
  year={1996},
  publisher={University of Chicago Press}
}

@article{connell1978diversity,
  title={Diversity in tropical rain forests and coral reefs: high diversity of trees and corals is maintained only in a nonequilibrium state.},
  author={Connell, Joseph H},
  journal={Science},
  volume={199},
  number={4335},
  pages={1302--1310},
  year={1978},
  publisher={American Association for the Advancement of Science}
}

@article{chesson1994multispecies,
  title={Multispecies competition in variable environments},
  author={Chesson, Peter},
  journal={Theoretical population biology},
  volume={45},
  number={3},
  pages={227--276},
  year={1994},
  publisher={Elsevier}
}

@article{d2008biodiversity,
  title={Biodiversity enhancement induced by environmental noise},
  author={D’Odorico, Paolo and Laio, Francesco and Ridolfi, Luca and Lerdau, Manuel T},
  journal={Journal of Theoretical Biology},
  volume={255},
  number={3},
  pages={332--337},
  year={2008},
  publisher={Elsevier}
}

@article{burkart2023periodic,
  title={Periodic temporal environmental variations induce coexistence in resource competition models},
  author={Burkart, Tom and Willeke, Jan and Frey, Erwin},
  journal={Physical Review E},
  volume={108},
  number={3},
  pages={034404},
  year={2023},
  publisher={APS}
}

@article{thebault2010stability,
  title={Stability of ecological communities and the architecture of mutualistic and trophic networks},
  author={Th{\'e}bault, Elisa and Fontaine, Colin},
  journal={Science},
  volume={329},
  number={5993},
  pages={853--856},
  year={2010},
  publisher={American Association for the Advancement of Science}
}

@article{mougi2012diversity,
  title={Diversity of interaction types and ecological community stability},
  author={Mougi, Akihiko and Kondoh, Michio},
  journal={Science},
  volume={337},
  number={6092},
  pages={349--351},
  year={2012},
  publisher={American Association for the Advancement of Science}
}

@article{bairey2016high,
  title={High-order species interactions shape ecosystem diversity},
  author={Bairey, Eyal and Kelsic, Eric D and Kishony, Roy},
  journal={Nature communications},
  volume={7},
  number={1},
  pages={12285},
  year={2016},
  publisher={Nature Publishing Group UK London}
}

@article{ser2018ubiquitous,
  title={Ubiquitous abundance distribution of non-dominant plankton across the global ocean},
  author={Ser-Giacomi, Enrico and Zinger, Lucie and Malviya, Shruti and De Vargas, Colomban and Karsenti, Eric and Bowler, Chris and De Monte, Silvia},
  journal={Nature ecology \& evolution},
  volume={2},
  number={8},
  pages={1243--1249},
  year={2018},
  publisher={Nature Publishing Group UK London}
}

@article{locey2016scaling,
  title={Scaling laws predict global microbial diversity},
  author={Locey, Kenneth J and Lennon, Jay T},
  journal={Proceedings of the National Academy of Sciences},
  volume={113},
  number={21},
  pages={5970--5975},
  year={2016},
  publisher={National Academy of Sciences}
}

@article{ferraro2026will,
  title={Will a time-varying complex system be stable?},
  author={Ferraro, Francesco and Grilletta, Christian and Maritan, Amos and Suweis, Samir and Azaele, Sandro},
  journal={arXiv preprint arXiv:2603.28464},
  year={2026}
}

@article{henderson2026fluctuating,
  title={Fluctuating environments are sufficient to drive substantial variability in species abundance across locations},
  author={Henderson, James FD and Tiffeau-Mayer, Andreas},
  journal={arXiv preprint arXiv:2603.02403},
  year={2026}
}

@article{calvo2026microbiome,
  title={Microbiome association diversity reflects proximity to the edge of instability},
  author={Calvo, Rub{\'e}n and Roig, Adri{\'a}n and L{\'o}pez, Roberto Corral and Camacho-Mateu, Jos{\'e} and Cuesta, Jos{\'e} A and Mu{\~n}oz, Miguel A},
  journal={arXiv preprint arXiv:2601.22918},
  year={2026}
}

@article{de2024many,
  title={Many-species ecological fluctuations as a jump process from the brink of extinction},
  author={de Pirey, Thibaut Arnoulx and Bunin, Guy},
  journal={Physical Review X},
  volume={14},
  number={1},
  pages={011037},
  year={2024},
  publisher={APS}
}

@article{levin1974dispersion,
  title={Dispersion and population interactions},
  author={Levin, Simon A},
  journal={The American Naturalist},
  volume={108},
  number={960},
  pages={207--228},
  year={1974},
  publisher={University of Chicago Press}
}

@article{macarthur1964competition,
  title={Competition, habitat selection, and character displacement in a patchy environment},
  author={MacArthur, Robert and Levins, Richard},
  journal={Proceedings of the National Academy of Sciences},
  volume={51},
  number={6},
  pages={1207--1210},
  year={1964}
}

@article{chesson1985coexistence,
  title={Coexistence of competitors in spatially and temporally varying environments: a look at the combined effects of different sorts of variability},
  author={Chesson, Peter L},
  journal={Theoretical Population Biology},
  volume={28},
  number={3},
  pages={263--287},
  year={1985},
  publisher={Elsevier}
}

@article{kalyuzhny2015neutral,
  title={A neutral theory with environmental stochasticity explains static and dynamic properties of ecological communities},
  author={Kalyuzhny, Michael and Kadmon, Ronen and Shnerb, Nadav M},
  journal={Ecology letters},
  volume={18},
  number={6},
  pages={572--580},
  year={2015},
  publisher={Wiley Online Library}
}

@article{loreau2008species,
  title={Species synchrony and its drivers: neutral and nonneutral community dynamics in fluctuating environments},
  author={Loreau, Michel and de Mazancourt, Claire},
  journal={The American Naturalist},
  volume={172},
  number={2},
  pages={E48--E66},
  year={2008},
  publisher={The University of Chicago Press}
}

@article{pande2022temporal,
  title={How temporal environmental stochasticity affects species richness: Destabilization, neutralization and the storage effect},
  author={Pande, Jayant and Shnerb, Nadav M},
  journal={Journal of Theoretical Biology},
  volume={539},
  pages={111053},
  year={2022},
  publisher={Elsevier}
}

@article{munoz1998nonlinear,
  title={On nonlinear diffusion with multiplicative noise},
  author={Munoz, Miguel A and Hwa, Terence},
  journal={EPL (Europhysics Letters)},
  volume={41},
  number={2},
  pages={147--152},
  year={1998}
}

@article{chesson2000general,
  title={General theory of competitive coexistence in spatially-varying environments},
  author={Chesson, Peter},
  journal={Theoretical population biology},
  volume={58},
  number={3},
  pages={211--237},
  year={2000},
  publisher={Elsevier}
}

@incollection{hairston1996overlapping,
  author    = {Hairston, Nelson G., Jr. and Ellner, Stephen P. and Kearns, Colleen M.},
  title     = {Overlapping Generations: The Storage Effect and the Maintenance of Biotic Diversity},
  booktitle = {Population Dynamics in Ecological Space and Time},
  editor    = {Rhodes, O. E., Jr. and Chesser, R. K. and Smith, M. H.},
  chapter   = {4},
  pages     = {109--145},
  year      = {1996},
  publisher = {University of Chicago Press},
  address   = {Chicago}
}

@article{garcia2022well,
  title={Well-mixed Lotka-Volterra model with random strongly competitive interactions},
  author={Garcia Lorenzana, Giulia and Altieri, Ada},
  journal={Physical Review E},
  volume={105},
  number={2},
  pages={024307},
  year={2022},
  publisher={APS}
}

@article{shmida1985biological,
  title={Biological determinants of species diversity},
  author={Shmida, AVI and Wilson, Mark V},
  journal={Journal of biogeography},
  pages={1--20},
  year={1985},
  publisher={JSTOR}
}

@article{leibold2004metacommunity,
  title={The metacommunity concept: a framework for multi-scale community ecology},
  author={Leibold, Mathew A and Holyoak, Marcel and Mouquet, Nicolas and Amarasekare, Priyanga and Chase, Jonathan M and Hoopes, Martha F and Holt, Robert D and Shurin, Jonathan B and Law, Richard and Tilman, David and others},
  journal={Ecology letters},
  volume={7},
  number={7},
  pages={601--613},
  year={2004},
  publisher={Wiley Online Library}
}

@article{mazzarisi2024complexity,
  title={Complexity-stability relationships in competitive disordered dynamical systems},
  author={Mazzarisi, Onofrio and Smerlak, Matteo},
  journal={Physical Review E},
  volume={110},
  number={5},
  pages={054403},
  year={2024},
  publisher={APS}
}

@article{chesson1981environmental,
  title={Environmental variability promotes coexistence in lottery competitive systems},
  author={Chesson, Peter L and Warner, Robert R},
  journal={The American Naturalist},
  volume={117},
  number={6},
  pages={923--943},
  year={1981},
  publisher={University of Chicago Press}
}

@article{picoche2019self,
  author  = {Picoche, Coralie and Barraquand, Fr{\'e}d{\'e}ric},
  title   = {How self-regulation, the storage effect, and their interaction contribute to coexistence in stochastic and seasonal environments},
  journal = {Theoretical Ecology},
  year    = {2019},
  volume  = {12},
  pages   = {489--500},
  doi     = {10.1007/s12080-019-0420-9}
}

@article{johnson2023coexistence,
  author  = {Johnson, Evan C. and Hastings, Alan},
  title   = {Coexistence in spatiotemporally fluctuating environments},
  journal = {Theoretical Ecology},
  year    = {2023},
  volume  = {16},
  pages   = {59--92},
  doi     = {10.1007/s12080-022-00549-7}
}

@article{seidelmann2025species,
  author  = {Seidelmann, Thomas and Mostaghim, Sanaz},
  title   = {Species coexistence as an emergent effect of interacting mechanisms},
  journal = {Theoretical Population Biology},
  year    = {2025},
  volume  = {162},
  pages   = {13--21},
  doi     = {10.1016/j.tpb.2024.12.005}
}

@article{tu1997systems,
  title={Systems with multiplicative noise: critical behavior from KPZ equation and numerics},
  author={Tu, Yuhai and Grinstein, G and Munoz, MA},
  journal={Physical review letters},
  volume={78},
  number={2},
  pages={274},
  year={1997},
  publisher={APS}
}

\end{document}


\title{Supplementary Information for ``Spatiotemporal noise stabilizes unbounded diversity in strongly-competitive communities"}
\author{Amer Al-Hiyasat}
\thanks{These authors contributed equally to this work.}
\author{Daniel W. Swartz}
\thanks{These authors contributed equally to this work.}
\author{Jeff Gore}
\author{Mehran Kardar}
\affiliation{Department of Physics, Massachusetts Institute of Technology, Cambridge, Massachusetts 02139, USA}
\date{\today}

\maketitle

\tableofcontents{}

\section{The finite-$P$ noise-driven extinction phase} \label{sec:noise-drivenExtinction}
Figure~\ref{fig:critical_behavior}b in the main text reports the long-time survival fraction as a function of noise strength $T$ and dispersal rate $D$, obtained from numerical simulations with large but finite $P$. The results show three possible phases as $T$ is increased, characterized by different scalings for the long-time number of survivors $S^\ast$:
\begin{enumerate}
    \item \textit{Competitive exclusion} ($T<T_c$): $S^\ast \in \mathcal{O}(1) \Rightarrow \phi \in \mathcal{O}(S^{-1})$.
    \item \textit{Noise-stabilized coexistence} ($T_c<T<T_\mathrm{ex}$): $S^\ast = S \Rightarrow \phi = 1$.
    \item \textit{Noise-driven extinction} ($T>T_\mathrm{ex}$): $S^\ast = 0 \Rightarrow \phi = 0$.
\end{enumerate}
Here, we show that phase (3) is not a true thermodynamic phase because $T_\mathrm{ex}$ diverges logarithmically in $P$, so that the $P \rightarrow \infty$ model displays only phases (1) and (2).

Since there are no survivors in phase (3), the extinction transition can be understood as a single-species problem \footnote{Concretely, we imagine extinctions happening sequentially, and consider the dynamics of the final extinction event in which there is only one remaining species.}. The dynamics then read
\begin{equation} \label{eq:singleSpecies} \dot{N} = rN(1-N) + D(\langle N \rangle - N) + \sqrt{2T} N \eta(t).
\end{equation}
The quantity $\langle N \rangle$ is defined as the patch average, which for finite $P$ is a fluctuating quantity. Extinction occurs when $\langle N \rangle(t)$ hits the cutoff $N_c \ll 1$.

\subsection{Single patch ($P=1$)}
The finite-$P$ transition to phase (3) is easiest to understand analytically on a single patch ($P=1$), where $\langle N \rangle = N$, so there is no dispersal term. Changing to log-abundances $z\equiv \log N$, the dynamics read
\begin{equation}\dot{z} = r - T - r e^{z} + \sqrt{2T} \eta(t),\end{equation}
where the $-T$ term arises from Ito's lemma. This is a Brownian motion with drift $r-T$ and an upper exponential wall at $z \simeq 0$. Given infinite time, $z(t)$ will hit the cutoff at $\log N_c \ll 0$ with probability 1. However, the time to extinction, $\tau_\mathrm{ex}$, scales differently with $N_c$ depending on the value of $T$. When $T<r$, the Brownian motion has positive drift, and the upper wall is important. The extinction time can be estimated by treating the exponential wall as a reflecting boundary at $z=0$. It then follows from standard results~\cite{gardiner2004handbook} that the $\tau_\mathrm{ex}$ will, for $N_c \ll 1$, scale as
\begin{equation} \tau_\mathrm{ex} \sim \frac{T}{(r-T)^2} N_c^{-(r-T)/T}\end{equation}
When $T>r$, on the other hand, the drift is negative and the upper wall is irrelevant, so the extinction time scales as
\begin{equation}\tau_\mathrm{ex} \sim \frac{|\log N_c|}{T-r}\end{equation}
We thus observe a transition at $T_\mathrm{ex} \equiv r$: For $T<T_\mathrm{ex}$, the extinction time diverges as a negative power of $N_c$, whereas for $T > T_\mathrm{ex}$ it is only logarithmic in $N_c$. Since the values of $N_c$ considered in this work are exponentially small, extinction is only numerically observed in the $T>T_\mathrm{ex}$ regime.

\subsection{Infinitely many patches ($P\rightarrow \infty$)}
To understand why adding patches delays the onset of noise-driven extinction, we first consider the infinite patch limit $P\rightarrow \infty$, where $\langle N \rangle$ becomes deterministic and equal to the ensemble average of $N$. It can then be shown analytically that, in absence of the cutoff ($N_c = 0$), Eq.~\eqref{eq:singleSpecies} has a $\langle N \rangle > 0$ steady-state irrespective of the value of $T$: To do this, we fix $\langle N \rangle$ and solve for the  stationary probability density $p(N)$ corresponding to Eq.~\eqref{eq:singleSpecies} . The mean abundance is then obtained self-consistently as $\langle N \rangle = \int dN N p(N)$. This is detailed in a prior work by some of us (Ref. \cite{ottino2020population}, see also Sec. below), where it is shown that a $\langle N \rangle > 0$ solution always exists. The log-abundance dynamics on a particular patch can then be written
\begin{equation}\dot{z} = -U'(z) + \sqrt{2T} \eta(t), \qquad U(z) \equiv -(r-T-D)z + r e^z + D \langle N \rangle e^{-z}.\end{equation}
Adding an extinction cutoff $N_c \ll \langle N \rangle$ will not affect this steady state on relevant time scales: on a particular patch, the time for $z(t)$ to hit the cutoff can be estimated through an Arrhenius-type argument and diverges exponentially as $\tau \sim e^{\beta U(\log N_c)}$. The time for $\langle N \rangle$ itself to reach the cutoff is then at least as large as this, so that
\begin{equation} \label{eq:extinctionTimeExp} \tau_\mathrm{ex} \gtrsim \exp\left(\frac{D \langle N \rangle}{T N_c}\right)
\end{equation}
The $\langle N \rangle > 0$ stationary solution, obtained with $N_c = 0$, is thus valid for $0<N_c \ll 1$, except on exponentially large time scales. We conclude that the noise-driven extinction phase does not survive the $P\rightarrow \infty$ limit.

\subsection{Large but finite networks ($1 \ll P \ll \infty$)}
\begin{figure}
    \centering
    \includegraphics[]{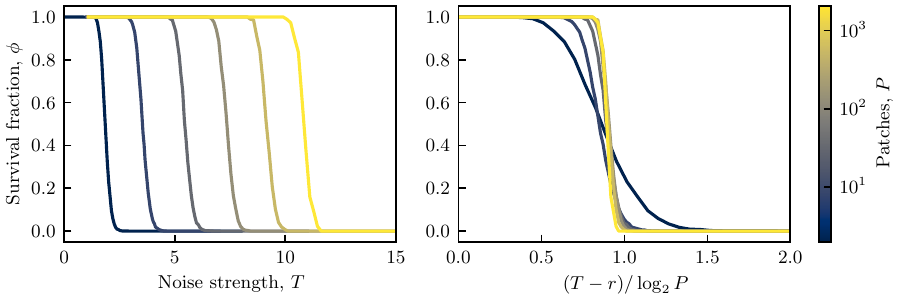}
    \caption{\textbf{Noise-driven extinction phase at finite $P$.} Left: Survival fraction as a function of noise strength for different $P$, showing extinction above a critical $T_\mathrm{ex}$. Right: data collapse when plotting $\phi$ as a function of $(T-r)/ \log P$, implying $T_\mathrm{ex} \sim \log P$. Species are noninteracting. $D=r=1, N_c = 10^{-15}$. }
    \label{fig:S1}
\end{figure}

For $1<P<\infty$, analytical treatment is more difficult. We instead report numerical simulations of Eq.~\eqref{eq:singleSpecies} for different $P$ in Fig.~\ref{fig:S1}. We observe an extinction transition from $\phi=1$ to $\phi = 0$ at a critical noise strength $T_\mathrm{ex}$. The curves can be collapsed by plotting as a function of $(T-r)/\log P$, implying a critical noise which diverges logarithmically in $P$:
\begin{equation}T_\mathrm{ex} = r + c \log P\end{equation}
for some $c(D) >0$ with dimensions of inverse time. To prevent extinction when $T>r$ thus requires the number of patches to be exponentially large in $T-r$.

\section{Species abundance distribution and Taylor's law} \label{sec:SADTaylor}
\subsection{Power law distribution over patches}
Standard treatments of the gLV equations using Dynamical Mean-Field Theory (DMFT) typically start by defining a stochastic process corresponding to the ``competitive load",
\begin{equation}\xi_i(t) \equiv \sum_{j \neq i} A_{ij} N_j(t)\end{equation}
and then fixing the correlators $\langle \xi(t) \rangle$ and $\langle \xi(t) \xi(t') \rangle$ by imposing self-consistency. We will show here that in the coexistence phase and for $D\gg r$, the self-consistent solution for $\xi$ has negligible fluctuations over time or between species. To do so, we replace $\xi$ with its stationary mean
\begin{equation}\xi_i \simeq \langle \xi_i \rangle = \sum_{j\neq i} A_{ij} \langle N_j\rangle,\end{equation}
and then show that $\langle \xi_i^2 \rangle_c \simeq 0$ self-consistently. Dropping the species index, our dynamics follow
\begin{equation} \label{eq:langevin}
\dot{N} = N(1 - \langle \xi \rangle - N) + D(\langle N \rangle - N) + \sqrt{2T} N \eta(t),
\end{equation}
where $r$ has been set to unity by rescaling time, so that $D$ and $T$ now represent the ratios $D/r$ and $T/r$, respectively. The corresponding Fokker-Planck equation reads:
\begin{equation} \label{eq:FPE} \partial_t p(N) = -\partial_N \{[N(1 - \langle \xi \rangle-N)+D(\langle N\rangle - N)]p(N)\} + T\partial_N^2 \left[N^2 p(N)\right]\,.\end{equation}
In the adiabatic limit $T,D \gg 1$, the mean abundance $\langle N \rangle$ may be held quasistatically fixed. All higher moments relax quickly to values determined by the stationary solution to Eq.~\eqref{eq:FPE}, which can be verified to take the truncated power law form~\cite{ottino2020population, swartz2022seascape, mallmin2025fluctuating}:
\begin{equation} \label{eq:statsol}
p(N) = \frac{1}{Z} e^{-\beta N} e^{-\beta D\langle N\rangle/N} N^{-\gamma},
\end{equation}
where
\begin{equation} \label{eq:gammadef} \beta \equiv \frac{1}{T}, \qquad \gamma \equiv 2 + \beta(D+\langle \xi \rangle-1), \qquad Z \equiv 2 \left[D \langle N\rangle\right]^{\frac{1-\gamma}{2}} K_{\gamma-1}\left(2\beta \sqrt{D \langle N \rangle}\right), \end{equation}
and $K_n$ denotes the $n$th modified Bessel function of the second kind.
The mean must satisfy the self-consistency relation $\langle N \rangle = \int_{0}^\infty z p(z) dz$, which upon evaluating the integral reads
\begin{equation} \label{eq:selfcon} \langle N \rangle = \frac{\sqrt{D \langle N \rangle} K_{\gamma-2}\left(2\beta \sqrt{D \langle N \rangle}\right)}{K_{\gamma-1}\left(2\beta \sqrt{D \langle N \rangle}\right)}.
\end{equation}
We will use this condition to determine $\langle \xi \rangle$. In the coexistence phase, we know from numerics that $\langle N \rangle \in \mathcal{O}(1/S)$. This motivates a small $\langle N \rangle$ expansion of Eq.~\eqref{eq:selfcon}, for which we rely on the series~\cite{ottino2020population}:
\begin{equation} \label{eq:besselExpansion} \frac{K_\nu(2x)}{K_{\nu+1}(2x)} = \frac{x}{\nu} \left[1+\frac{x^2}{\nu (1-\nu)} + \frac{\Gamma(-\nu)}{\Gamma(\nu)} x^{2\nu} + \mathcal{O}(x^4, x^{2\nu +2})\right].\end{equation}
Defining $x \equiv \beta \sqrt{D \langle N \rangle}$, we have from Eq.~\eqref{eq:selfcon},
\begin{align}
    x &= \beta D \frac{K_{\gamma-2}(2x)}{K_{\gamma-1}(2x)} \nonumber \\
    &\simeq \frac{\beta D}{\gamma - 2}x \left(1+ \frac{x^2}{(\gamma-2)(3-\gamma)}+ \frac{\Gamma(2-\gamma)}{\Gamma(\gamma-2)}x^{2\gamma-4}\right) \,.\label{eq:meanExpansion}
\end{align}
Using the definition of $\gamma$ in Eq.~\eqref{eq:gammadef}, we may now solve for $\langle \xi \rangle$ perturbatively in powers of $x$. For $\gamma > 2$, the terms in $x^2$ and $x^{2\gamma-4}$ are subleading. Self-consistency at leading order then requires $\gamma - 2 = \beta D$, from which we obtain
\begin{equation} \label{eq:xiSol}
    \langle \xi \rangle = 1 + o(1).
\end{equation}
Notably, at leading order, $\langle \xi \rangle$ does not depend on $x$, so that every species has the same value of $\langle \xi \rangle$ even if the mean abundances $\langle N_i \rangle$ are different (i.e., the quenched fluctuations of $\xi(t)$ are vanishing). This does not hold at finite $S$: the leading correction to Eq.~\eqref{eq:xiSol} can be obtained from Eq.~\eqref{eq:meanExpansion} as
\begin{equation} \label{eq:selfOrg} \langle \xi \rangle - 1 = \frac{T}{1-\beta D}x^2  + D \frac{\Gamma(-\beta D)}{\Gamma(\beta D)} x^{2\beta D} + \mathcal{O}(x^4, x^{2+2\beta D}).\end{equation}
This scales as $S^{-1}$ for $T<D$ and as $S^{-\beta D}$ for $T>D$. Substituting back into Eq.~\eqref{eq:langevin} yields a dynamics with a renormalized growth rate $1-\langle \xi \rangle$ that vanishes with $S$, constituting a form of self-organized criticality \cite{de2025self, denk2022self}. For $S\rightarrow \infty$, setting $\langle \xi \rangle = 1$ shows that every species follows a power law distribution with the same exponent, but with a different lower cutoff set by the mean $\langle N_i \rangle$:
\begin{equation} \label{eq:statDist} p(N_i) =\frac{1}{Z} e^{-\beta N_i} e^{-\beta D \langle N_i \rangle/N_i} N_i^{-2-\beta D}.
\end{equation}
This may be interpreted as a distribution over patches at any point in time, or, if the mean $\langle N_i \rangle$ has reached its steady state, as a distribution over time on a single patch.

\subsection{Anomalous scaling of the second moment}
The second moment of the stationary distribution (Eq.~\ref{eq:statsol}) can be evaluated as
\begin{equation}\left \langle N^2 \right\rangle = D \langle N \rangle \frac{K_{\gamma-3}\left(2\beta \sqrt{D \langle N \rangle} \right)}{K_{\gamma-1}\left(2\beta \sqrt{D \langle N \rangle}\right)}.\end{equation}
Rewriting $\langle N \rangle$ using Eq.~\eqref{eq:selfcon} puts this into the form of Eq.~\eqref{eq:besselExpansion}, allowing for a small $x \equiv \beta \sqrt{D \langle N \rangle}$ expansion. We find:
\begin{equation}\frac{\left \langle N^2 \right\rangle}{\langle N \rangle} \simeq \frac{T x^2}{\beta D - 1} \left[1 + \frac{x^2}{(\beta D - 1)(2-\beta D)} + \frac{\Gamma(1-\beta D)}{\Gamma(\beta D - 1)} x^{2(\beta D - 1)} \right],\end{equation}
where we have used the solution $\gamma = 2+ \beta D$ (Eq.~\ref{eq:xiSol}). The term in $x^2$ is always subleading, whereas the term in $x^{2(\beta D - 1)}$ is subleading for $T<D$ and dominant for $T>D$. We thus have a transition at $T = D$~\cite{ottino2020population, swartz2022seascape}:
\begin{equation} \label{eq:secmom}
\left \langle N^2 \right\rangle =\begin{cases}
\frac{D}{D-T} \, \langle N \rangle^2,  &(T<D), \\
 a(T,D) \, \langle N \rangle^{1+\beta D}, &(T>D),
\end{cases}
\end{equation}
where
\begin{equation}a(T,D) \equiv \frac{\Gamma(1-\beta D)}{\Gamma(\beta D)} (\beta D)^{\beta D} T^{1-\beta D}. \end{equation}
The result for $T<D$ recovers the adiabatic large-$D$ solution from the main text (Eq.~\ref{eq:glvclosure}). The $T>D$ case is new and shows an anomalous power law scaling of the second moment in the coexistence phase, with a tunable exponent $1+\beta D$ \cite{ottino2020population, swartz2022seascape}. This is an instance of Taylor's law of fluctuation scaling~\cite{taylor1961aggregation, eisler2008fluctuation}, thus providing a mechanistic origin for this empirical macroecological law.

With these results in hand, we revisit the self-consistency of the approximation $\xi \simeq \langle \xi_i \rangle$. We have
\begin{equation}
\left \langle \xi_{i}^2 \right \rangle = \sum_{k, j} A_{ij} A_{ik} \langle N_j N_k\rangle
\end{equation}
In the $D \gg 1$ regime, we may invoke Eq.~\eqref{eq:offDiagClosure} of the main text, yielding
\begin{align*}
    \left \langle \xi_i^2 \right \rangle &= \sum_j A_{ij}^2 \left \langle N_j^2\right \rangle + \sum_{k,j} A_{ij} A_{ik} \left \langle N_j \right \rangle \left \langle N_k \right \rangle - \sum_{j} A_{ij}^2 \left \langle N_j \right \rangle^2\\
    &= \mathcal{O}\left(S^{-\beta D}\right) + \left \langle \xi_i \right \rangle^2 + \mathcal{O}\left(S^{-1}\right).
\end{align*}
As $S \rightarrow \infty$, we have $\left \langle \xi_i^2 \right \rangle = \left \langle \xi_i \right \rangle^2$. The quantity $\xi_i(t)$ thus has neither temporal nor interspecific fluctuations.

We stress that the derivation above relies on being in the coexistence phase, where the mean abundances are $\mathcal{O}(1/S)$. The results are thus guaranteed to hold only above the coexistence threshold $T_c$, which will be determined in the next section both for constant and disordered interaction matrices.

\section{The $\theta$-logistic generalized Lotka-Volterra model} \label{sec:theta}
The results of Section~\ref{sec:SADTaylor}, together with the $D\gg r$ factorization approximation $\langle N_i N_{j\neq i} \rangle = \langle N_i \rangle \langle N_j \rangle$ of Eq.~\eqref{eq:offDiagClosure}, imply the patch-averaged abundances to evolve according to the $\theta$-gLV equations given in Eq.~\eqref{eq:effectiveDynamics} of the main text, repeated below:

\begin{equation} \label{eq:effectiveDynamics-SI}
    \langle \dot{N_i} \rangle = r \langle N_i \rangle \bigg[1-  \left(\frac{\langle N_i \rangle}{K}\right)^{\theta} - \sum_{j\neq i} A_{ij} \langle N_j \rangle \bigg],
\end{equation}
where $\theta$ and $K$ depend on the noise and dispersal rate as
\begin{equation}\theta(D,T) = \begin{cases}
    1, & T<D,\\
    \beta D, & T>D,
        \end{cases}
\qquad \qquad K(D,T) = \begin{cases}
    1-T/D, & T<D,\\
    \frac{1}{\theta}\left[\frac{\Gamma(\theta)}{\Gamma(1-\theta)}\right]^{1/\theta}\left(\frac{T}{r}\right)^{1-1/\theta}, & T>D.
\end{cases}\end{equation}
In this section, we study the fixed points of the $\theta$-gLV model and determine their stability, allowing us to explain and characterize the novel coexistence phase.
The analysis presented here follows similar lines to that in Ref.~\cite{hatton2024diversity}, though our focus on the $\theta>0$ regime changes several of the results and allows for further analytical progress in the presence of disordered interactions. For brevity, we rescale the patch-averaged abundance, the interaction matrix, and the time variable as
\begin{equation}M_i \equiv \langle N_i \rangle/K, \qquad A\rightarrow A K, \qquad t \rightarrow t/r\,.\end{equation}
The dynamics then read
\begin{equation} \label{eq:thetalog} \dot{M_i} = M_i\bigg[1-M_i^\theta - \sum_{j\neq i} A_{ij} M_j\bigg].\end{equation}
For $T<D$, we have $\theta = 1$, corresponding to the standard gLV model. Within the coexistence phase $T> T_c$, we instead have $\theta = \beta D$.

\subsection{Constant interactions}
We first treat the case $A_{ij} = \mu$, which can be solved directly through a large-$S$ perturbation theory. We recast the equation in terms of the variables
\begin{equation}y_i \equiv \varepsilon^{-1/\theta} M_i, \qquad \varepsilon \equiv \left[\mu(S-1)\right]^{-\theta}. \end{equation}
The quantity $\varepsilon \ll 1$ will prove to be a natural expansion parameter, with $y_i \in \mathcal{O}(\varepsilon^0)$ at the coexisting fixed point. The rescaled equations read:
\begin{equation} \dot{y}_i = y_i\bigg[1-\varepsilon y_i^\theta - \mu \varepsilon^{1/\theta} \sum_{j\neq i} y_j\bigg]. \label{eq:rescaledEqn} \end{equation}
We look for a fixed point in which all species coexist. By symmetry, all abundances must be equal at this point: $y_1 = \dots = y_S \equiv y > 0$. The fixed point condition then reads:
\begin{equation} \label{eq:fpEqn} 1-y-\varepsilon y^\theta = 0.
\end{equation}
Though this admits no closed-form solution, we can derive an expansion for $y$ in powers of $\varepsilon$. The zeroth-order solution is immediately read off as $y = 1+\mathcal{O}(\varepsilon)$, but this order will turn out insufficient to determine stability. To obtain the leading corrections, we write $y =1+\delta$ and substitute into~\eqref{eq:fpEqn}, yielding
\begin{equation}\delta + \varepsilon(1+\delta)^\theta = 0.\end{equation}
We now invoke the binomial expansion $(1+\delta)^\theta = 1 + \theta \delta + \frac{1}{2}\theta(\theta-1)\delta^2 + \mathcal{O}(\delta^3)$, and further expand $\delta = \sum_{n=1}^\infty c_n \varepsilon^n$. The $c_n$ can then be determined order by order in $\varepsilon$, yielding the following expression for the fixed-point abundance:
\begin{equation}\label{eq:fpvalue}y = 1-\varepsilon+\theta \varepsilon^2 + \mathcal{O}(\varepsilon^3).\end{equation}
Having located the fixed point, we now determine its stability. The Jacobian of Eq.~\eqref{eq:rescaledEqn} can be computed as
\begin{equation}J_{k \ell}(\{y_i\}) \equiv \partial_\ell \dot{y}_k= \delta_{k\ell}\bigg[1-\varepsilon(\theta+1)y_k^\theta + \mu \varepsilon^{1/\theta} y_k - \mu \varepsilon^{1/\theta} \sum_{j\neq k} y_j\bigg] - \mu \varepsilon^{1/\theta} y_k.\end{equation}
At the fixed point $y_i = y$, this equals
\begin{equation}J^\ast_{k\ell} = \delta_{k\ell}\left[1-(\theta+1)(1-y) - (1-\mu \varepsilon^{1/\theta})y\right] - \mu \varepsilon^{1/\theta} y,\end{equation}
where we have used Eq.~\eqref{eq:fpEqn} to replace $y^\theta$ with $(1-y)/\varepsilon$. The matrix $J^\ast$ has one eigenvector $(1,\dots, 1)^T$ with eigenvalue $\lambda_1$, and $S-1$ degenerate eigenvalues $\lambda_2$ whose eigenvectors are orthogonal to $(1,\dots, 1)^T$. It can be verified that $\lambda_1 = \lambda_2 - \mu \varepsilon^{1/\theta} S y$, and
\begin{equation}\lambda_2 = 1-(\theta+1)(1-y)-(1-\mu \varepsilon^{1/\theta})y.\end{equation}
Since $\lambda_1 < \lambda_2$, stability relies on the negativity of $\lambda_2$. Using the small $\varepsilon$ expansion of $y$ (Eq.~\ref{eq:fpvalue}), we find
\begin{equation}\lambda_2(\varepsilon) = -\theta \varepsilon + \mu \varepsilon^{1/\theta} + \mathcal{O}(\varepsilon^2, \varepsilon^{1+1/\theta}).\end{equation}
In the special case $\theta = 1$, we find $\lambda_2(\varepsilon) = (\mu -1) \varepsilon + \mathcal{O}(\varepsilon^2)$, which yields the usual gLV condition for stability, $\mu < 1$. If $\theta < 1$, the term in $\varepsilon^{1/\theta}$ is subleading, and so $\lambda_2$ is always negative for sufficiently small $\varepsilon$ (large $S$). We conclude that, for constant interactions $A_{ij} = \mu$, the coexisting fixed point is asymptotically stable whenever $\theta < 1$, as reported in Ref.~\cite{hatton2024diversity}. Using $\theta=\beta D$, this implies that the critical noise strength for coexistence is $T_c = D$.

\subsection{Disordered interactions}
We now treat the case of random heterogeneous interactions. The starting point is Eq.~\eqref{eq:thetalog} with quenched disorder in $A$, whose moments are specified as
\begin{equation}\overline{A_{ij}} = \mu, \qquad \overline{(A_{ij} A_{k\ell})_c} = \sigma^2 \delta_{ik}\delta_{j\ell},\end{equation}
where $\mu, \sigma^2 > 0$ are kept $\mathcal{O}(S^0)$.
A fixed point for which $M_i > 0$ must satisfy
\begin{equation}1-M_i^\theta - \sum_{j\neq i} A_{ij} M_j = 0.\end{equation}
We now separate $A$ into its mean and fluctuations $A_{ij} = \mu + \sigma a_{ij}$, where the $a_{ij}$ are i.i.d. with zero mean and unit variance. We then have
\begin{equation} \label{eq:fpCondition} M_i^\theta = 1 - \mu S \overline{M} - \sigma \sum_{j\neq i} a_{ij} M_j,
\end{equation}
where $\sum_{j \neq i} M_j$ has been replaced with $S \overline{M} \equiv \sum_j M_j$, which is valid for large $S$. The final term can be treated by the cavity method in the limit $S \rightarrow \infty$, which justifies replacing $\sum_{j\neq i} a_{ij} M_j$ with a Gaussian variable whose statistics are computed by treating the $M_j$ as fixed and independent of the $a_{ij}$:
\begin{equation} \label{eq:cavity}
\Big\langle \sum_{j \neq i} a_{ij} M_j \Big \rangle \simeq \sum_{j \neq i} \langle a_{ij} \rangle M_j = 0, \qquad \bigg\langle \Big(\sum_{j \neq i} a_{ij} M_j \Big)^2 \bigg \rangle \simeq \sum_{j,k \neq i} \langle a_{ij} a_{i k} \rangle M_j M_k  = \sum_{k \neq i} M_k^2 \equiv S \overline{M^2}.
\end{equation}
We note that, although standard derivations of the cavity method rely on the scaling of $\mu, \sigma^2$ as $S^{-1}$, the forms posited in Eq.~\eqref{eq:cavity} will prove self-consistent in the coexistence phase, owing to the $\mathcal{O}(S^{-1})$ scaling of $M_i$. Equations~\eqref{eq:fpCondition} and~\eqref{eq:cavity} together give
\begin{equation} \label{eq:cavityN^theta} M_{i}^\theta = 1-\mu S \overline{M} -  \sigma \sqrt{S \overline{M^2}} \,Z ,
\end{equation}
where $Z \sim \mathcal{N}(0,1)$ represents the quenched (interspecific) fluctuations. If, for a given realization of $Z$, the right hand side of Eq.~\eqref{eq:cavityN^theta} is negative, then there is no real and positive solution for $M_i$, and species $i$ cannot be extant at the fixed point. Allowing for the extinction solution $M_i = 0$, we define the distribution of a typical fixed-point abundance through the random variable
\begin{equation} \label{eq:cavityDist} M = \max\left(0, 1-\mu S \overline{M} -  \sigma \sqrt{S \overline{M^2}} \,Z\right)^{1/\theta}, \qquad Z \sim \mathcal{N}(0,1).
\end{equation}
$\nbar$ and $\nnbar$ are then determined by the self-consistency equations
\begin{equation} \label{eq:selfConsistencyRaw} \overline{M^n} = \frac{1}{\sqrt{2\pi \sigma^2 S \nnbar}} \int_0^\infty dx\, x^{n/\theta} e^{-\frac{(x-1+\mu S \nbar)^2}{2\sigma^2 S \nnbar}}, \qquad (n>0). \end{equation}
These equations may be solved numerically by recursion, as is usually done for the gLV case ($\theta = 1$) with interactions scaling as $\mu, \sigma^2 \propto S^{-1}$. However, we will show here that with interactions of order unity, the self-consistent problem can be solved analytically in closed form for large $S$, revealing a novel coexistence phase when $\theta < 1/2$.

For notational convenience, we let $m$ and $g^2$ denote the mean and variance of the shifted Gaussian appearing in Eq.~\eqref{eq:cavityDist}:
\begin{equation}m \equiv 1-\mu S \nbar, \qquad g^2 \equiv \sigma^2 S \nnbar,\end{equation}
so that we may write $M = \max(0, \zeta)^{1/\theta}$, with $\zeta \sim \mathcal{N}(m, g^2)$. For a given species to survive, we require $\zeta > 0$. The probability of this is
\begin{equation}\phi = \mathbb{P}[M>0] = \frac{1}{\sqrt{2\pi}} \int_{-m/g}^\infty dx e^{-x^2/2}.\end{equation}
Searching for a coexistence phase, we demand that $\lim_{S \rightarrow \infty} \phi = 1$. This requires $m/g \rightarrow +\infty$ as $S$ is made large, so that $\zeta$ is almost surely positive in this limit. We posit the divergence to occur with some positive exponent $\eta$:
\begin{equation} \label{eq:hetaDef} \frac{m}{g} \sim h S^\eta, \qquad \eta, h>0,
\end{equation}
where $h \in \bigO{1}$ is a prefactor that will be determined self-consistently. We furthermore define exponents describing the large-$S$ behavior of $\nbar$ and $\nnbar$:
\begin{equation}\nbar \sim \tnbar S^{-\alpha}, \qquad \nnbar \sim \tnnbar S^{-\gamma},\end{equation}
where $\tnbar$ and $\tnnbar$ are the respective $\bigO{1}$ parts. The program is now to substitute these scaling forms into the self-consistency equations and expand in $S$ to obtain identities relating the exponents $\alpha, \gamma, \eta$ and $\theta$.

We first rewrite the self-consistency equation~\eqref{eq:selfConsistencyRaw} through the change of variable $z \equiv (x - m)/g$:
\begin{align*}\overline{M^n} &= \frac{1}{\sqrt{2\pi}} \int_{-m/g}^\infty dz (m+ gz)^{n/\theta} e^{-z^2/2} \\
&= \frac{m^{n/\theta}}{\sqrt{2\pi}} \int_{-h S^\eta}^\infty dz \left(1+\frac{z}{h S^\eta}\right)^{n/\theta} e^{-z^2/2}.\end{align*}
To study the large-$S$ behavior, we expand the factor $\left(1+z h^{-1} S^{-\eta}\right)^{n/\theta}$ as a binomial series, reducing the integral to a sum over truncated Gaussian moments
\begin{equation}
\overline{M^n} = \frac{m^{n/\theta}}{\sqrt{2\pi}} \sum_{k=0}^\infty \binom{n/\theta}{k} h^{-k} S^{- k \eta} \int_{-h S^\eta}^{\infty}dz z^k e^{-z^2/2} \,.
\end{equation}
For large $S$, we may extend the lower integration bound to $-\infty$, incurring only an exponentially-small error. We then find, using the formula for Gaussian central moments,
\begin{align} \label{eq:selfConSeries} \overline{M^n} = m^{n/\theta} \sum_{k=0}^\infty \binom{n/\theta}{2k} h^{-2k} S^{-2k\eta} (2k-1)!! \quad+ \bigO{e^{-S^{2\eta}}}.
\end{align}
The leading term of this series suffices to set conditions on the scaling exponents. For $n=1$ and $n=2$, we find:
\begin{align}
    \tnbar &= m^{1/\theta} S^{\alpha} = \left(g h S^\eta\right)^{1/\theta} S^\alpha = \left[h^2 \sigma^2 \tnnbar\right]^{1/2\theta} S^{\frac{1-\gamma}{2\theta}+\frac{\eta}{\theta}+\alpha}, \label{eq:AsymSelfConNbar}\\
    \tnnbar &= m^{2/\theta} S^{\gamma} = (g h S^\eta)^{2/\theta}S^{\gamma} = \left[h^2 \sigma^2 \tnnbar\right]^{1/\theta} S^{\frac{1-\gamma}{\theta}+\frac{2\eta}{\theta}+\gamma}. \label{eq:AsymSelfConN2bar}
\end{align}
Demanding that the above remain $\bigO{1}$ as $S \rightarrow \infty$, we obtain the following identities:
\begin{align}
    1-\gamma+2\eta+2\theta \alpha &= 0, \label{eq:identity1}\\
    1+(\theta-1)\gamma + 2\eta &=0 \label{eq:identity2}.
\end{align}
Subtracting these equations gives
\begin{equation}\gamma = 2\alpha.\end{equation}
Using this relation and the definition in Eq.~\eqref{eq:hetaDef},
\begin{equation} \label{eq:mgScaling} h S^\eta \simeq \frac{m}{g} = \frac{1-\mu \tnbar S^{1-\alpha}}{\sigma \sqrt{\tnnbar} S^{\frac{1}{2}-\alpha}}.
\end{equation}
For the above to be positive, as required for coexistence, the numerator imposes that $\alpha \geq 1$. But if $\alpha > 1$, then $\eta = \alpha - 1/2$. Substituting this into Eq.~\eqref{eq:identity1} yields $\theta \alpha = 0$, which is a contradiction unless $\theta = 0$. We thus must have $\alpha = 1$, implying $\gamma = 2$, and, by Eqs.~\eqref{eq:identity1} and~\eqref{eq:identity2}, $\eta = \frac{1}{2} - \theta$. For this to be self consistent, Eq.~\eqref{eq:mgScaling} requires
\begin{equation}h S^{- \theta} \simeq \frac{1-\mu \tnbar}{\sigma \sqrt{\tnnbar}}.\end{equation}
But since $\tnbar$ and $\tnnbar$ are $\bigO{1}$ by construction, we must have
\begin{equation} \label{eq:uNofh} \mu \tnbar = 1 - h \sigma \sqrt{\tnnbar} S^{-\theta}.\end{equation}
Thus, $\tnbar = 1/\mu$ at leading order, but the $\bigO{S^{-\theta}}$ correction must be included for self-consistency. The leading-order result suffices to determine $h$: from Eqs.~\eqref{eq:AsymSelfConNbar} and ~\eqref{eq:AsymSelfConN2bar}, we have $\tnnbar = \tnbar^2$ at order $S^0$, so that
\begin{equation}\tnnbar = \mu^{-2} + o(1).\end{equation}
Equation~\eqref{eq:AsymSelfConNbar} then gives $\mu^{-1} = (h \sigma \mu^{-1})^{1/\theta}$, implying
\begin{equation}h=\frac{\mu^{1-\theta}}{\sigma}.\end{equation}
Using these results, Eq.~\eqref{eq:uNofh} becomes $\mu \tnbar = 1 - (\mu S)^{-\theta}$.

To summarize, we have solved the self-consistent cavity problem explicitly in the large-$S$ limit, finding
\begin{equation} \label{eq:summary} \nbar \simeq \frac{1}{\mu S}\left[1 - \frac{1}{(\mu S)^\theta} + o(S^{-\theta})\right], \qquad \nnbar \simeq \frac{1}{\mu^2 S^2} + o(S^{-2}).
\end{equation}
Recalling that $\eta=1/2-\theta>0$ was required for an asymptotic survival fraction of unity, we obtain the coexistence criterion:
\begin{equation} \theta < \frac{1}{2}.\end{equation}
Notice that, at leading order, we have $\nnbar = \nbar^2$, implying that disorder is irrelevant in the true $S \rightarrow \infty$ limit, and that all abundances converge to the mean $\nbar$ defined in Eq.~\eqref{eq:summary}.

\section{The small $D$ regime} \label{sec:smallD}
Our analytical theory of the noise-stabilized coexistence phase relies on the adiabatic limit $D \gg r$, where the factorization approximation $\langle N_i N_j \rangle \approx \langle N_i \rangle \langle N_j \rangle$ becomes valid (Fig.~\ref{fig:richards}). However, the coexistence phase is observed numerically over a broad range of $D$, including for $D \ll r$, as seen in Fig.~\ref{fig:critical_behavior}B of the main text. Here, we ask which of the conclusions of the analytical theory carry over to the $D \ll r$ regime.

\begin{figure}
    \centering
    \includegraphics[trim=8 0 7  0, clip]{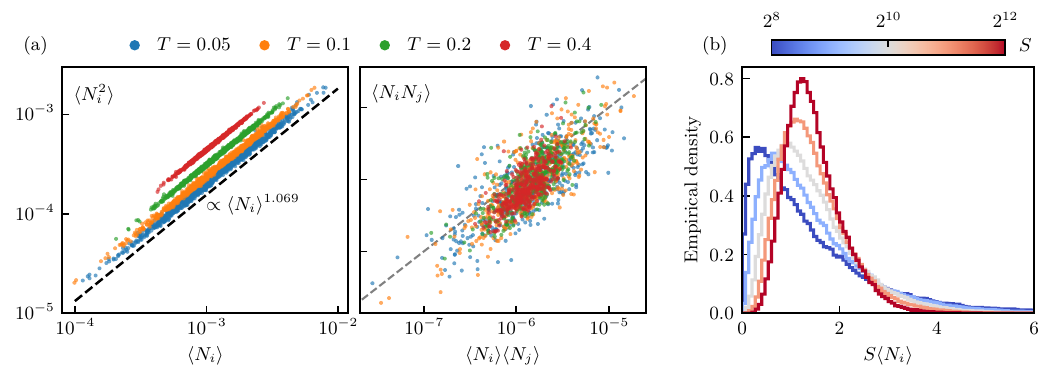}
    \caption{Scaling of moments in the $D\ll r$ regime ($D=10^{-4}r$). (a) Left: Dependence of the steady-state second moment $\langle N_i^2 \rangle$ on the mean $\langle N_i \rangle$ for different noise strengths, $T$. Each point is a different species from a single metacommunity simulation. Taylor's law scaling $\langle N_i^2 \rangle \propto \langle N_i \rangle^{1+\theta}$ is obeyed, with the same best-fit $\theta$ for all values of $T$ tested, $\theta = 0.069 \pm 0.001$. Right: Dependence of the two-species moment $\langle N_i N_j \rangle$ on the product $\langle N_i \rangle \langle N_j \rangle$. The large-$D$ factorization approximation $\langle N_i N_j \rangle =\langle N_i \rangle \langle N_j \rangle$ (gray dashed line) clearly breaks down. $S = 1024, P=2^{18}$. (b) Distribution of patch-averaged abundances in the coexistence phase for different $S$. Data is consistent with a narrowing of the distribution in the high diversity limit, as in the $\theta$-gLV model. $T=0.1, P=2^{16}$. In both panels, $D = 10^{-4}r$ and $A_{ij} \in \mathcal{N}(0.7, 0.2^2)$.}
    \label{fig:smallD}
\end{figure}

Figure~\ref{fig:smallD} reports simulations of large metacommunities with $D = 10^{-4} r$. As shown in Fig.~\ref{fig:smallD}a (right), the factorization approximation for $\langle N_i N_j \rangle$ clearly breaks down in this regime. However, Taylor's law scaling $\langle N_i^2 \rangle \propto \langle N_i \rangle^{1+\theta}$ is obeyed robustly, with an exponent that can be obtained by fitting.
For all tested noise strengths (ranging from $T = 0.05$ to $T=0.4$), the fits yield values of $\theta$ that are indistinguishable within error, giving an estimate of $\theta = 0.069 \pm 0.001$. The large-$D$ prediction is $\theta = \beta D$, which evaluates to $\theta = 0.002$ for $T= 0.05$ and $\theta =0.0025$ for $T = 0.4$. The prediction $\theta = \beta D$ thus appears to fail in the small-$D$ regime, though we note that for values of $\theta$ this small, finite-size effects scaling as $S^{-\theta}$ are difficult to control; for the data in Fig.~\ref{fig:smallD}, for example, we have $S^{-\theta} = 1024^{0.069} \approx 1.6$. Together, the results of Fig.~\ref{fig:smallD}a suggest that it is the Taylor's law scaling which is essential to the coexistence mechanism, not the two-species factorization approximation.

Without the factorization approximation, our analytical mapping of the dynamics of $\langle N_i(t) \rangle$ to the $\theta$-gLV model cannot be applied exactly. Nonetheless, several qualitative predictions of the $\theta$-gLV analysis of Sec.~\ref{sec:theta} apply here: Figure~\ref{fig:smallD}b shows the distribution of stationary patch-averaged abundances for different $S$. As in the $\theta$-gLV, we find that $\langle N_i \rangle$ scales as $1/S$. The distribution of $\langle N_i \rangle$ also appears to narrow for large $S$, as in Fig.~\ref{fig:thetamapping}C of the main text, though the rate at which this occurs is very slow,  leaving an appreciable width even at $S = 4096$. Once again, the smallness of $\theta$ in this regime makes finite-$S$ effects important.

\section{Fitness-density covariance} \label{sec:gdCovar}
To connect our work to Chesson's modern coexistence theory~\cite{chesson2000mechanisms, chesson2000general}, we consider here a particular decomposition of the dynamics of the patch-averaged abundances considered for a different model in Ref.~\cite{chesson2000general}. Let us denote by $f_{i\alpha}$ the local fitness (per capita growth rate) of species $i$ on patch $\alpha$,
\begin{equation} f_{i\alpha} \equiv 1 - N_{i\alpha} - \sum_{j\neq i} A_{ij} N_{j\alpha},\end{equation}
so that the dynamics read $\dot{N}_{i\alpha} = N_{i\alpha} f_{i\alpha} + D \left(\langle N_i \rangle - N_{i\alpha}\right) + \sqrt{2T}\eta_{i\alpha} $. Averaging over patches then allows us to decompose the dynamics of the mean exactly as
\begin{align} \label{eq:gdcovariance}
    \frac{\langle \dot{N}_{i}\rangle}{\langle N_{i}\rangle} = \frac{\langle N_{i} f_{i}\rangle}{\langle N_{i} \rangle} = \langle f_i \rangle +\mathrm{Cov}\left(\frac{N_i}{\langle N_i \rangle}, f_i\right).
\end{align}
The first term on the RHS is the mean fitness and the second is the \textit{fitness-density covariance}. For a ``resident" species, $\langle \dot{N}_i \rangle = 0$, and the two terms must cancel. For a rare invader to invade, we require the two terms to sum to a positive number during the course of invasion.

\begin{figure}
    \centering
    \includegraphics{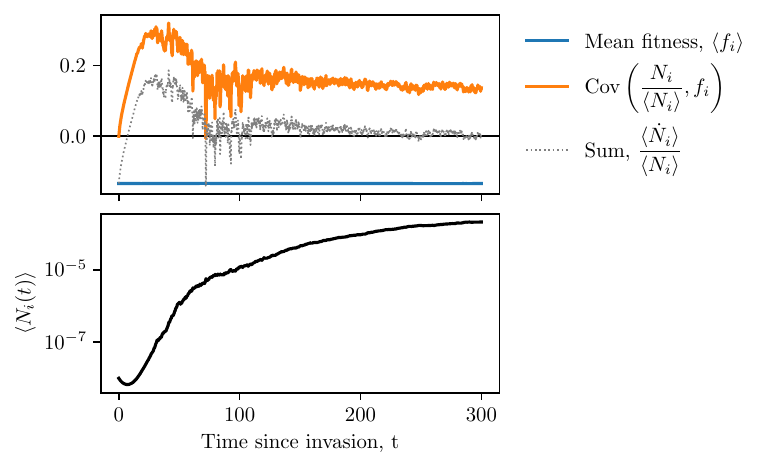}
    \caption{Fitness-density covariance of invading species. A steady-state metacommunity in the noise-stabilized coexistence phase is prepared, and an invader species $i$ is introduced uniformly in space at an initial abundance of $10^{-8}$. Upper panel plots the two contributions to the mean per capita growth rate of the invader identified in Eq.~\eqref{eq:gdcovariance}, as well as their sum (dotted gray line). Lower panel shows the mean abundance of the invader.
    $S = 256, P = 2^{21}, D = 10^{-3}, A_{ij} \sim \mathcal{N}(0.7, 0.2^2)$.}
    \label{fig:gdcov}
\end{figure}

To compare the contributions of the two terms above, we performed numerical simulations in which a coexisting metacommunity is prepared in the steady state, and an invader species is seeded at low abundance throughout space. For the example shown in Fig.~\ref{fig:gdcov}, the invader has a negative mean fitness throughout the invasion, and its abundance initially falls. During the invasion, however, positive fitness-density covariance develops, indicating that the species accumulates where its competitive fitness is instantaneously higher. This is sufficient to overcome the negative mean fitness and allow successful invasion. Eventually, the mean fitness and fitness-density covariance become equal and opposite, at which point the invader becomes a resident.

\section{Weak interactions} \label{sec:weakInteractions}
Throughout this work, we have emphasized that spatiotemporal noise can stabilize coexistence in spite of \textit{strong} competitive interactions; i.e. with $\mu$ and $\sigma^2$ of order unity, so that the noiseless system is subject to the diversity-stability paradox. In this section, we consider weak interactions, in which the strength of interactions is made to vanish with the number of species as $\mu, \sigma^2 \propto 1/S$. Under this scaling, a finite survival fraction is obtained even without noise, with coexisting abundances of order unity (and thus a total biomass which diverges with $S$)~\cite{bunin2017ecological}.

\subsection{Survival fraction}
\begin{figure}
    \centering
    \includegraphics{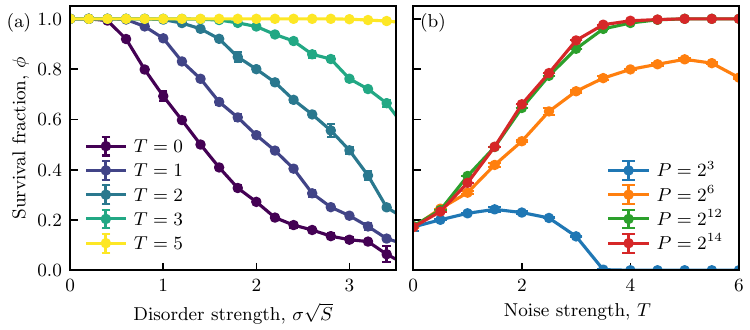}
    \caption{Spatiotemporal noise increases survival fraction, $\phi$, in weakly-interacting metacommunities. (a) $\phi$ as a function of disorder strength $\sigma \sqrt{S}$ for different noise strengths $T$. $P = 2^{14}, S = 2048.$ (b) $\phi$ as a function of noise strength $T$ for different patch counts $P$ at fixed $\sigma = 2.5/\sqrt{S}$, showing that the transition does not sharpen in the $P\rightarrow \infty$ limit, unlike the strongly-interacting case. $S = 1024$. The $P = 8$ curve shows the onset of noise-induced extinction at large $T$, as discussed in Sec.~\ref{sec:noise-drivenExtinction}.  In both panels, $\mu = 10/S, D = 2$, and diversity is measured at $t_\mathrm{max} = 10^4$.}
    \label{fig:weakSurvival}
\end{figure}

Our analytical theory of the strongly-interacting case relies on expansions in $\langle N_i \rangle \propto 1/S$ and thus does not immediately extend to weak interactions, where $\langle N_i \rangle \in \mathcal{O}(1)$. We instead rely on numerical simulations: Figure.~\ref{fig:weakSurvival}(a) shows that, in a large weakly-interacting metacommunity, adding spatiotemporal noise increases the survival fraction beyond its $T=0$ value. This is true for any degree of interaction disorder,  regardless of whether the system is in the unique fixed point or multiple attractors phase~\cite{bunin2017ecological}. Figure.~\ref{fig:weakSurvival}(b) plots the survival fraction as a function of the noise strength at fixed interaction disorder. Unlike in the strongly interacting case, where a sharp, discontinuous transition from $\phi = 0$ to $\phi = 1$ occurs at $T = T_c(D)$ in the $P \to \infty$ limit (Fig.~\ref{fig:critical_behavior}), the data here are consistent with a crossover from $\phi(T=0) \in (0, 1)$ to $\phi(T \to \infty) = 1$ that remains continuous even for large $P$.~\footnote{Strictly, we have not ruled out the possibility of a continuous transition to $\phi = 1$ at some $T_c$; the numerical data are insufficient to distinguish this from a smooth crossover}.

\begin{figure}
    \centering
    \includegraphics[width=\textwidth]{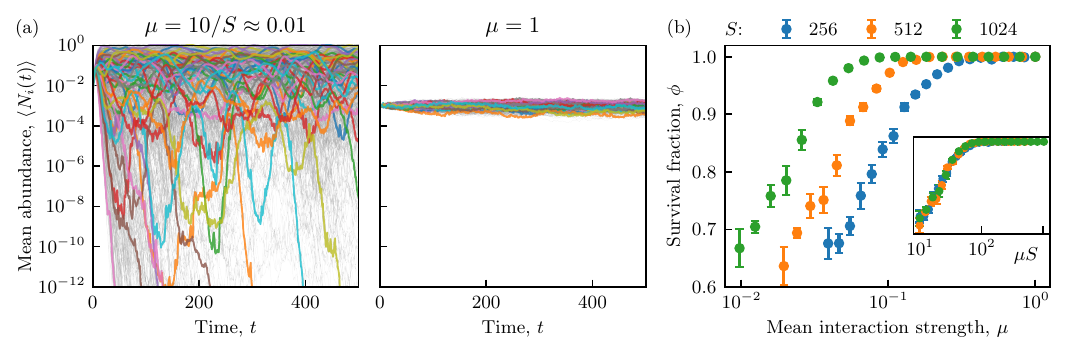}
    \caption{Increasing competition can increase the survival fraction in weakly-interacting metacommunities with spatiotemporal noise. (a) Patch-averaged abundance trajectories with $\mu \in \mathcal{O}(1/S)$ (left) or $\mu \in \mathcal{O}(1)$ (right). Forty species are highlighted in a background of $S = 1024$. In the weakly-interacting case (left), several extinctions are observed, whereas with strong interactions (right) all species persist. (b) Survival fraction as a function of mean interaction strength, $\mu$, for different $S$. Inset: data collapse when plotted as a function of $\mu S$, showing that $\phi = 1$ when $\mu S \gg 1$.
    In both panels, $P=2^{13}, T = 0.5, D = 0.1$ and interaction disorder scales weakly $\sigma = 3/\sqrt{S}$.}
    \label{fig:phiOfmu}
\end{figure}

A surprising consequence of this is revealed by considering a weakly-interacting system in which $T$ exceeds the strongly-interacting transition threshold $T_c(D)$, as in Fig.~\ref{fig:phiOfmu}: In this case, increasing $\mu$ from $\mathcal{O}(1/S)$ to $\mathcal{O}(1)$ to cross over from weak to strong interactions \textit{increases} the survival fraction to $\phi=1$ (Fig.~\ref{fig:phiOfmu}b). This increase in survival fraction is accompanied by a decrease in total biomass from $\mathcal{O}(S)$ to $\mathcal{O}(1)$.
We thus conclude that in metacommunities with environmental noise, stronger competition can, paradoxically, protect against competitive exclusion by suppressing total biomass.

\subsection{Species abundance distribution} \label{sec:weakSAD}
\begin{figure}
    \centering
    \includegraphics{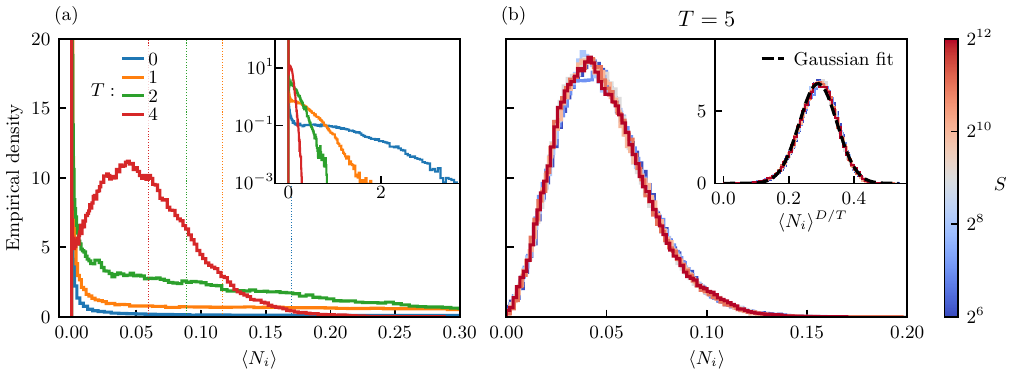}
    \caption{Distribution of patch-averaged abundances in a $P = 2^{14}$ metacommunity with weak interactions $\mu = 10/S, \sigma = 2.5/\sqrt{S}$, and $D=2.0$. (a) Distribution of $\langle N_i \rangle$ for different $T$ at fixed $S= 2048$. Vertical dotted lines indicate the mean $\overline{\langle N \rangle}$. Inset: Same data with a logarithmic vertical axis and extended horizontal axis, showing that for extant species, typical $\langle N_i \rangle$ become smaller when $T$ is increased. (b) Distribution of $\langle N_i \rangle$ for different $S$  ranging from 64 to 4096, at fixed $T = 5$. Inset: distribution of $\langle N_i \rangle^{\theta=D/T}$, with a Gaussian fit indicated.}
    \label{fig:weakSAD}
\end{figure}

We next consider the empirical distribution of patch-averaged abundances $\langle N_i\rangle$. In the strongly-interacting case, this was found to narrow to a sharp peak at $1/\mu S$ as $S\rightarrow \infty$, showing that interaction disorder is effectively neutralized in the high diversity limit and that species thus become indistinguishable. In weakly interacting systems, however, since the survival fraction is less than unity, interaction disorder remains relevant and selects which species go extinct. In Fig.~\ref{fig:weakSAD}(a), the we show the distribution of $\langle N_i \rangle$ for different $T$. There is a delta mass at $\langle N_i \rangle = 0$ corresponding to extinct species. The distribution of extant species narrows as $T$ is increased, and the typical nonzero $\langle N_i \rangle$ become smaller (Fig.~\ref{fig:weakSAD}a, inset). This means that, although increasing $T$ increases the survival fraction, the total biomass $S\overline{\langle N \rangle}$ is decreased (Fig.~\ref{fig:weakSAD}a, dotted lines).

 Extant species remain distinguishable in the high diversity limit: In Fig.~\ref{fig:weakSAD}(b), the distribution is plotted for different $S$, with $T$ chosen large enough that $\phi$ is indistinguishable from unity for the values of $S$ tested. There is clearly a large-$S$ limiting form, with a mean of order unity and a width that does not vanish with $S$. This can be understood from the fact that, with weak interactions, fluctuations in interactions become larger than the mean in the large-$S$ limit, as $\sigma/\mu \sim \sqrt{S}$, and typical abundances remain $\mathcal{O}(1)$, so that a term like $\sum_{j\neq i} A_{ij} N_i N_j$ has mean and fluctuations of order unity.

The general form of this limiting distribution cannot be obtained exactly from our analytical theory, as the mapping to the $\theta$-gLV model relies on $\langle N_i \rangle \propto 1/S \ll 1$, whereas $\langle N_i \rangle \in \mathcal{O}(1)$ for weak interactions. Nonetheless, if most $\langle N_i \rangle$ are small, the theory becomes a good approximation. This is the case in Fig.~\ref{fig:weakSAD}(b), where $\overline{\langle N \rangle} \approx 0.06$ because $T$ is large. We then expect a typical patch-averaged abundance to be distributed according to $\langle N \rangle = \mathrm{min}(0,\zeta)^{1/\theta}$, with $\zeta$ a Gaussian whose mean and variance are determined self-consistently. The large-$D$ theory in turn predicts $\theta = D/T$, so that $\langle N_i \rangle^{D/T}$ is expected to follow a truncated Gaussian distribution. This is verified in the inset to Fig.~\ref{fig:weakSAD}(b), where $\phi$ is close to unity and it is seen that $\langle N_i \rangle^{D/T}$ indeed fits well to a Gaussian.

\section{Random growth rates} \label{sec:randomGrowth}
In our noise-stabilized coexistence phase, all species coexist, and interaction disorder is asymptotically irrelevant in the high diversity limit. Different species thus become indistinguishable as $S\rightarrow \infty$. However, interaction disorder is not the sole driver of interspecific variation--- species may also differ in their intrinsic growth rates or dispersal rates, and may be subject to different levels of environmental noise.

As a first step in this direction, we present, in this section, numerical results on the effects of growth rate disorder. We consider the following generalization of our model:
\begin{equation} \label{eq:randomGrowthRate}
    \dot{N}_{i\alpha} = r N_{i\alpha} \left[g_i - N_{i\alpha} - \sum_{j\neq i} A_{ij} N_{j\alpha}\right] + D\left(\langle N_i \rangle - N_{i\alpha}\right) + \sqrt{2T} N_{i \alpha} \eta_{i \alpha}(t),
\end{equation}
where $g_i$ is a quenched dimensionless growth rate, which differs between species but is uniform in space and time. The dimensionful prefactor, $r$, remains to set the overall ecological time scale.

\begin{figure}
    \centering
    \includegraphics{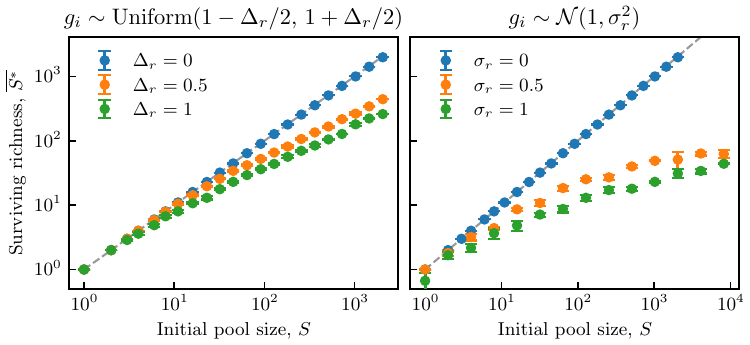}
    \caption{Effect of growth rate disorder on the number of survivors. Numerical simulations of Eq.~\eqref{eq:randomGrowthRate} were performed for either a uniform distribution of growth rates (left), or a Gaussian distribution (right).
    Surviving richness is plotted as a function of the initial number of species for different strengths of growth rate disorder. In absence of disorder, the number of survivors grows as $S^\ast \propto S$. For bounded disorder (left), it appears to grow as $S^\ast \propto S^\eta$ for some $\eta<1$, whereas for unbounded disorder (right) $S^\ast$ saturates. In both panels, $P = 2^{16}, D = 2, T=5, A_{ij}\sim \mathcal{N}(2,1)$.}
    \label{fig:growthRateSurvival}
\end{figure}
We first ask whether the unbounded coexistence phase persists. Preliminary numerical results suggest that the behavior depends qualitatively on whether $g_i$ is drawn from a bounded or an unbounded distribution (for example, a uniform or a Gaussian distribution, respectively). In both cases, Fig.~\eqref{fig:growthRateSurvival} shows that the full coexistence of all species with $\phi=1$ does not survive the addition of growth rate disorder, and that extinction is possible even in the strong noise regime when the $g_i$ are different. However, in the unbounded $g_i$ case, the number of survivors $S^\ast$ appears to saturate as a function of the initial pool size $S$, whereas for bounded $g_i$, data are more consistent with a scaling
\begin{equation} \label{eq:etascaling}
    S^\ast \propto S^\eta,
\end{equation}
for some $\eta<1$. This would suggest that, in the presence of bounded growth rate disorder, spatiotemporal noise still stabilizes unbounded diversity, but with a survival fraction that vanishes asymptotically as $\phi \propto S^{\eta-1}$. Further numerical work will be necessary to establish Eq.~\eqref{eq:etascaling} conclusively and determine the parameter dependence of $\eta$.

\begin{figure}
    \centering
    \includegraphics{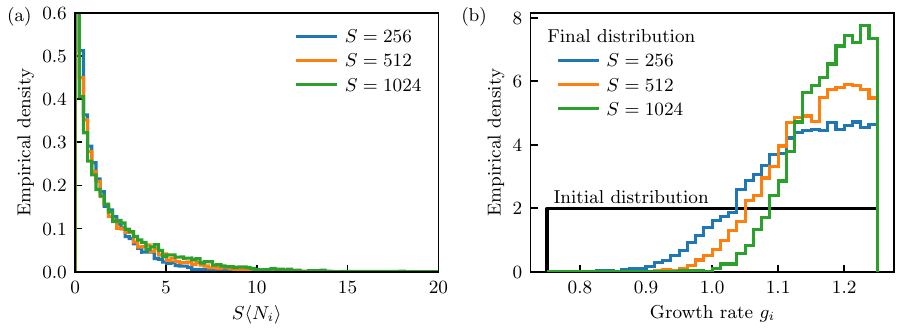}
    \caption{Abundance distribution in the presence of growth rate disorder $g_i \sim \mathrm{Uniform}(0.75,1.25)$. (a) Stationary distribution of patch-averaged abundances for different initial pool size $S$. The peak at $\langle N_i \rangle = 0$ corresponding to extinct species is truncated.  (b) Growth rate distribution of surviving species for different initial pool size. Initial uniform growth rate distribution is indicated in black. Data demonstrates the selection of fast growing species.
    In both panels, $P = 2^{16}, D = 2, T=5, A_{ij}\sim \mathcal{N}(2,1)$.}
    \label{fig:randomRSAD}
\end{figure}

Growth rate disorder qualitatively changes the distribution of patch-averaged abundances into the form shown in Fig.~\eqref{fig:randomRSAD}a. Unlike in the constant $g_i$ case, this distribution does not become peaked around its mean for large $S$. In Fig.~\eqref{fig:randomRSAD}b, the distribution of growth rates among surviving species is plotted, showing that the ecological dynamics select for faster-growing species, and that this selection is more stringent for larger $S$.

\section{Correlated noise} \label{sec:CorNoise}
\subsection{Temporal correlations} \label{sec:temporalNoise}
In this work, we have focused for simplicity on white environmental noise, which is formally valid when the correlation time of environmental fluctuations is short compared to the generation time and the inverse dispersal rate. Real environmental fluctuations, however, might occur on time scales comparable to generation time, which invites us to consider the effects of temporal correlations in environmental noise.

\begin{figure}
    \centering
    \includegraphics{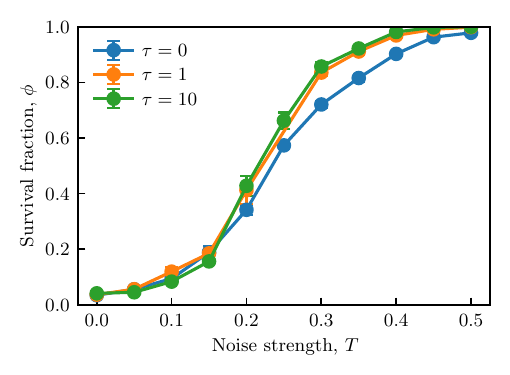}
    \caption{Coexistence transition in metacommunities with temporally-correlated noise of correlation time $\tau$ (Eq.~\ref{eq:ouNoise}). The stationary survival fraction changes from $\phi\simeq 0$ to $\phi=1$ as $T$ is increased, and increases only slightly when $\tau>0$. $P=2^{14}, S = 256, D = 0.1, A_{ij} \sim \mathcal{N}(0.7, 0.2^2)$.}
    \label{fig:temporalNoise}
\end{figure}

The self-organized criticality which accompanies coexistence (Eq.~\ref{eq:selfOrg}) implies a corresponding critical slowing down: since the renormalized mass vanishes as $S^{-\theta}$, the characteristic time scale diverges as $S^\theta$. This suggests that any finite correlation time in the noise will be irrelevant in the large-$S$ limit. To verify this, we consider an exponentially correlated noise:
\begin{equation} \label{eq:ouNoise} \langle \eta_{i \alpha}(t) \eta_{j \beta}(t') \rangle = \frac{\delta_{ij}\delta_{\alpha \beta}}{2\tau} e^{|t'-t|/\tau}, \end{equation}
where $\tau$ is the correlation time, and the factor of $1/\tau$ ensures a proper white noise limit as $\tau \rightarrow 0$ (though this limit corresponds to a Stratonovich rather than an Ito process). Fig.~\ref{fig:temporalNoise} shows that the coexistence transition persists in the presence of temporal correlations $\tau>0$, and its onset is not appreciably modified by variations in $\tau$.

\subsection{Spatial correlations} \label{sec:spatialCor}
We next consider the role of spatial correlations in environmental noise. For the fully-connected spatial network considered here, this can be studied by modifying the correlation function of the noise into
\begin{equation} \label{eq:spatialCor}
\langle\eta_{i\alpha}(t)\eta_{j\beta}(t')\rangle
=\delta_{ij}\left[(1-\rho)\delta_{\alpha\beta}+\rho\right] \delta(t-t'),
\end{equation}
so that $\rho = 0$ corresponds to the independent patch case studied throughout this work, and $\rho = 1$ corresponds to having all patches experiencing exactly the same noise. This is sampled numerically by writing $\eta_{i\alpha}$ as the linear combination of an i.i.d. white noise, with weight $\sqrt{1-\rho}$, and a single global noise variable with weight $\sqrt{\rho}$.

\begin{figure}
    \centering
    \includegraphics{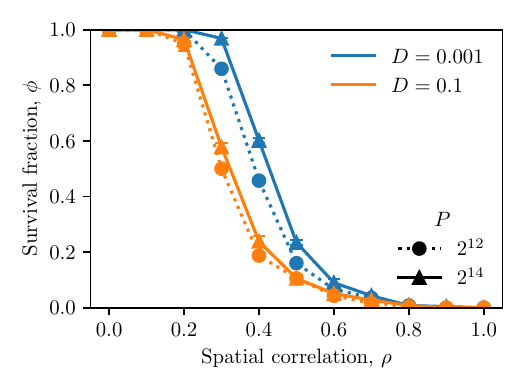}
    \caption{Effect of spatial correlations in environmental noise on the survival fraction in the coexistence phase. Correlations are implemented according to Eq.~\eqref{eq:spatialCor}. The dispersal rate, $D$, and number of patches, $P$, are indicated. $S=128, T = 1.0, A_{ij} \sim \mathcal{N}(0.7, 0.2^2)$.}
    \label{fig:spatialNoise}
\end{figure}

Figure.~\ref{fig:spatialNoise} plots the survival fraction, $\phi$, as a function of the spatial correlation, $\rho$, for two values of $D$. We find that the $\phi = 1$ coexistence phase is robust to small positive values of $\rho$, but that $\phi$ collapses down to zero if $\rho$ is made large enough. Increasing the number of patches does not eliminate this behavior, but lower dispersal rates allow the community to sustain larger $\rho$. Further numerical work is needed to distinguish a sharp transition at a critical $\rho_c$ from a smooth crossover and, in the former case, to determine how $\rho_c$ depends on $D$ and $T$.

We note that, in finite-dimensional spatial systems, such as the one- and two-dimensional lattices considered in the next section, spatial correlations are more naturally introduced via a finite correlation length $\ell$ over which noise decorrelates \big[e.g., $\langle \eta_{i}(\mathbf{x}, t) \eta_{j}(\mathbf{x}', t)\rangle \propto \delta_{ij} \delta(t-t') e^{-\vert{}\mathbf{x} - \mathbf{x}'\vert{}/\ell}$\big]. By analogy with the temporally correlated case in Sec.~\ref{sec:temporalNoise}, if the self-organized criticality of the fully connected model extends to finite dimensions, the correlation lengths of the abundance fields will diverge with a power of $S$~\cite{munoz1998nonlinear}. This should render any finite $\ell$ irrelevant in the $S, P \rightarrow \infty$ limit, up to renormalization of the parameters.

\section{Beyond fully connected, infinite networks}
In the main text and in the above, we showed that spatiotemporal noise can stabilize the coexistence of arbitrarily-many species on a fully-connected and infinite spatial network. Real environments, however, are finite-dimensional spaces of finite extent. To address this, this section presents preliminary numerical results on finite fully-connected networks as well as spatial lattices in one and two dimensions.

\subsection{Fully connected networks with finite $P$}
\begin{figure}
    \centering
    \includegraphics[]{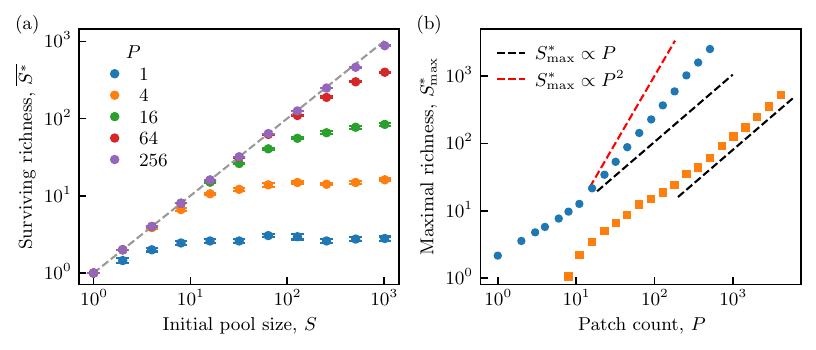}
    \caption{Surviving richness is limited only by metacommunity size. (a) Average surviving richness $\overline{S^\ast}(S, P)$ as a function of initial species pool size for different patch counts $P$. $D = 0.1, T = 0.5, \mu = 0.7, \sigma = 0.2$. Richness is sampled at $t = 10^4$.
    (b) Estimated maximal surviving richness, $S_\mathrm{max}^\ast \equiv \lim_{S\rightarrow \infty} \overline{S^\ast}(S, P)$, as a function of the number of patches for two different parameter sets. Richness is sampled at a time $t_\mathrm{max} \sqrt{P}$, chosen based on separate measurements on the convergence time of $S^\ast(P)$ (not shown). Blue circles correspond to: $D=10^{-4}, T = 0.5, \mu = 2\sigma = 0.8, t_\mathrm{max} = 10^5$. Orange squares correspond to $D=1, T = 3, \mu = 2\sigma = 2, t_\mathrm{max} = 640$. In both cases, the extinction cutoff is $N_c = 10^{-15}$ and the initial pool size is $S = 4096$. Data represents the average of many replicates. }
    \label{fig:finitePScaling}
\end{figure}

Figure~\ref{fig:fig1}D of the main text shows that the long-time number of survivors $S^\ast(S, P)$ is equal to the pool size $S$ when $P\rightarrow \infty$. When the number of patches is finite, we show in Fig.~\ref{fig:finitePScaling}a  that $S^\ast(S, P)$ grows initially with $P$ but ultimately saturates to a maximal value $S^\ast_{\mathrm{max}}(P)$, which we define as
\begin{equation}
    S^\ast_{\mathrm{max}}(P) \equiv \lim_{S\rightarrow \infty} \overline{S^\ast(S, P)}
\end{equation}
This maximal sustainable richness grows without bound as $P$ is increased, showing that diversity is limited only by system size. To estimate the dependence on $P$, we performed numerical simulations for different $P$, starting with a large pool size $S \gg S^\ast_\mathrm{max}(P)$. Our results confirm that $S^{\ast}_\mathrm{max}(P)$ grows with $P$, but suggest that the large-$P$ asymptotic scaling is nonuniversal. This is illustrated in Fig.~\ref{fig:finitePScaling}b for two representative parameter sets: in one case, with $D,T \in \mathcal{O}(1)$ (orange squares in Fig.~\ref{fig:finitePScaling}b), our results are consistent with a linear scaling $S^\ast_\mathrm{max}(P) \propto P$, suggesting a maximal richness per patch. In another case, with $D \ll T < 1$ (blue circles in Fig.~\ref{fig:finitePScaling}b), we find $S^\ast_\mathrm{max}(P) \propto P^{\alpha}$ with $1 < \alpha < 2$. The latter case is interesting because it suggests that coexistence can be stabilized with $S \gg P$ for sufficiently large $S, P$. Further numerical analysis is necessary to determine the detailed dependence of $S^\ast$ on $P$, which connects to the broader empirical literature on species-area relationships~\cite{rosenzweig1995species}.

\subsection{One- and two-dimensional lattices} \label{sec:finiteDim}
Next, we consider the case where the network of patches has the connectivity of a hyper-cubic lattice in $d$ dimensions. Equation~\eqref{eq:abundance_full} of the main text then becomes:
\begin{equation} \label{eq:squareLattice}
\dot{N}_{i, \mathbf{r}} =
    r N_{i, \mathbf{r}} \bigg[ 1 - N_{i, \mathbf{r}} - \sum_{j\neq i} A_{ij}N_{j, \mathbf{r}} \bigg] + \frac{D}{2d}  \sum_{\beta = 1}^d \left( N_{i, \mathbf{r} + \mathbf{e}^{(\beta)}} + N_{i, \mathbf{r} - {\mathbf{e}}^{(\beta)}} - 2 N_{i, \mathbf{r}} \right) + \sqrt{2T} N_{i, \mathbf{r}} \eta_{i , \mathbf{r}}(t),
    \end{equation}
where $\mathbf{r} \in \mathbb{Z}^d$ is the spatial index and ${\mathbf{e}}^{(\beta)}$ is the unit vector in direction $\beta$, with entries ${e}^{(\beta)}_\alpha = \delta_{\alpha \beta}$.
This may be interpreted as the discretization of a continuous space $\mathbb{R}^d$ by defining an appropriate dimensionful diffusivity $\tilde{D} \equiv a^2 D/2 d$, and noise $\tilde{T} \equiv a^d T$, where $a$ is the lattice spacing.

\begin{figure}
    \centering
    \includegraphics[]{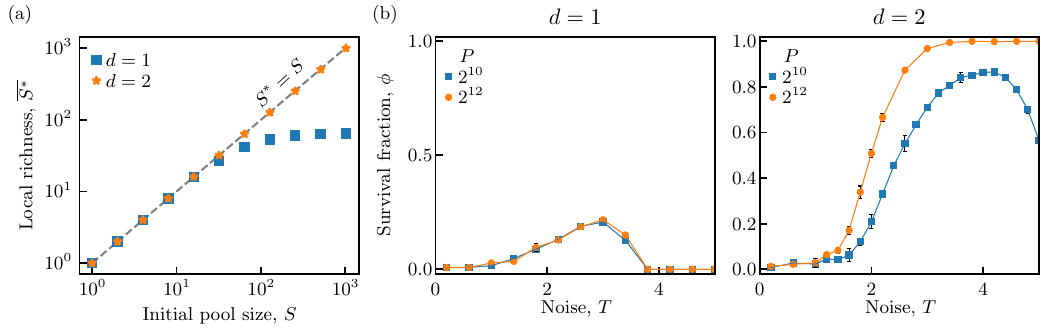}
    \caption{One- and two-dimensional lattice patch networks. (a) Surviving richness as a function of initial pool size in spatial dimensions $d=1$ or $d=2$. The line $S^\ast = S$ is indicated. In $d=2$, all species persist up to the maximum measured pool size $S=1024$, whereas for $d=1$ the richness saturates. (b) Survival fraction as a function of noise strength $T$ for $d=1,2$ and for the indicated patch numbers $P$ (corresponding to system volume). The initial increase with $T$ represents noise-stabilized coexistence, whereas the fall for large $T$ represents noise-driven extinction. In $d=1$, full coexistence is never achieved, and increasing $P$ does not delay noise-driven extinction. In contrast, the $d=2$ case resembles the fully-connected results (Fig.~\ref{fig:critical_behavior} of the main text), showing a well-defined coexistence transition and no noise-driven extinction for sufficiently large $P$.}
    \label{fig:finitedim}
\end{figure}

We performed numerical simulations of Eq.~\eqref{eq:squareLattice} in $d=1$ and $d=2$. Figure~\ref{fig:finitedim}(a) reports the surviving richness, $S^\ast$, as a function of the initial pool size $S$, in analogy to Fig.~\ref{fig:fig1}(c) of the main text. The results for $d=2$ resemble the fully-connected case, with $S^\ast = S$, even for $S$ as large as $10^3$. In $d=1$, however, the surviving richness saturates to a value of the order $10^2$. Figure~\ref{fig:finitedim}(b) reports the survival fraction as a function of noise strength: once again, the $d=2$ case resembles the fully-connected model, displaying, for sufficiently large $P$, a transition from $\phi \simeq 0$ to $\phi = 1$ at a threshold noise strength $T_c$. In $d=1$, this transition is not observed, and the data furthermore suggest that the large-$T$ noise-driven extinction phase is not abrogated by increasing $P$.

More detailed analysis will be necessary to conclude on the fate of the coexistence phase in finite dimensions; in particular, measurements of the $\theta$ exponent will be necessary to determine the stability of coexistence in the infinite-$S$ limit (see Ref.~\cite{swartz2022seascape}, where such a measurement is reported for a single species in $d=1$). Present data is consistent with the hypothesis that the coexistence phase persists in $d=2$, but not $d=1$. In the latter case, spatiotemporal noise increases richness but cannot stabilize truly extensive diversity.
We plan to pursue this question in a future work.

\section{Numerical details}

\subsection{Insensitivity of coexistence to the extinction cutoff} \label{sec:cutoff}
As stated in the main text, demographic extinction was implemented numerically using a hard cutoff, $N_c$, below which the local abundance is set to zero. Here, we verify that the precise value of the cutoff is irrelevant to the survival fraction in the coexistence phase, so long as $N_c$ is well separated from the $\mathcal{O}(D/S)$ migration floor.
\begin{figure}
    \centering
    \includegraphics{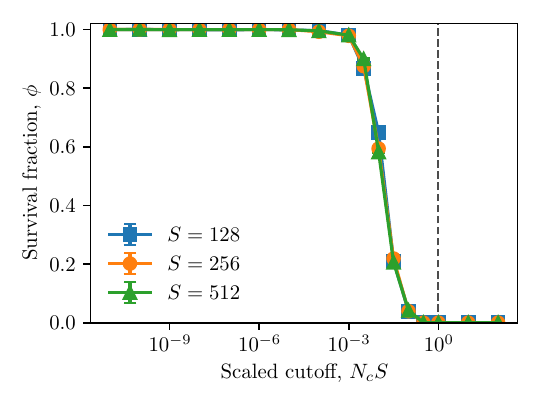}
    \caption{Dependence of survival fraction on the extinction cutoff in the noise-stabilized coexistence phase. Error bars represent standard error over independent replicates. $P=2^{14}, T = 0.5, D = 0.1, A_{ij}\sim\mathcal{N}(0.7, 0.2^2)$.}
    \label{fig:cutoff}
\end{figure}

Figure~\ref{fig:cutoff} plots the survival fraction as a function of the scaled cutoff $N_c S$ for different $S$, showing collapse of the data onto a single curve. The same parameters as in Fig.~\ref{fig:fig1} of the main text are used. For $N_c \ll 1/S$, the survival fraction is unity, whereas when $N_c \sim 1/S$, $\phi$ collapses to zero. All simulations in this work use $N_c \ll 1/S$.

\subsection{Simulation methods}
Simulations of the stochastic metacommunity dynamics were performed using the Milstein method, though the standard Euler-Maruyama method achieves comparable performance for the observables reported in this work. Since our simulations involve a large number of fields, $S$, over a large number of spatial points, $P$, they are uniquely amenable to massively-parallel GPU acceleration. This allows for simulations where the number of variables is on the order of a billion (for example, Fig.~\ref{fig:thetamapping} of the main text includes simulations with $SP = 2^{32} \approx 4\times10^9$).

For the simulation with Ornstein-Uhlenbeck (OU) noise in Sec.~\ref{sec:temporalNoise}, the noise at time $t+\Delta t$ is sampled exactly according to
\begin{equation}\eta(t+\Delta t) = e^{-\Delta t/\tau} \eta(t) + \sqrt{\frac{1-e^{-2\Delta t/\tau}}{\tau}} Z_t,\end{equation}
where $Z_t \sim \mathcal{N}(0,1)$. The multiplicative noise update to the abundance is implemented using the trapezoidal rule $N(t+\Delta t ) = N(t)(1+x+x^2/2)$, where $x= \sqrt{2T} \Delta t[\eta(t) + \eta(t+\Delta t)]/2$, which is valid so long as $\Delta t \ll \tau$. For $\tau = 0$, we use $x = \sqrt{2T \Delta t} Z_t$, for which the trapezoidal update yields an appropriate Stratonovich dynamics.

All numerical methods were implemented in the Julia programming language using CUDA.jl~\cite{besard2018juliagpu} for GPU programming, and run on a combination of Nvidia L40S, H100, and A30 GPUs. We are grateful to the MIT Office of Research Computing and Data (ORCD) and the SubMIT facility~\cite{bendavid2025submitphysicsanalysisfacility} at MIT Physics for providing GPU resources.

\subsection{Simulation parameter values for figures in the main text}
All measurements of species richness or survival fraction represent quasi-stationary values, defined by sampling at a time $t_\mathrm{max}$ such that $S^\ast(t_\mathrm{max}/2)$ differs from $S^\ast(t_\mathrm{max})$ by less than 10\%. For large $S, P$ and small $N_c$, our measurements are consistent with the following empirical scaling form for the relaxation time of $S(t)$:
\begin{equation}
    t_\mathrm{max} \propto \sqrt{P} \log\frac{S}{N_c}
\end{equation}
This is well-separated from the noise-driven extinction time, which is exponential in $1/N_c$ (Eq.~\ref{eq:extinctionTimeExp}), showing that the quasi-stationary regime is well-defined for small $N_c$ (that is, large populations).

Most simulation parameter values are listed in the Figure captions; below, we list remaining ones that were left out of the captions for brevity. For all simulations, and integration time step of $\Delta t \leq 0.01$ was used.
\begin{itemize}
    \item \textbf{Figure 1:}
        \begin{enumerate}[label=(\alph*)]
            \item[(c)] $t_\mathrm{max} = 8000$.
        \end{enumerate}
    \item \textbf{Figure 2:}
        \begin{enumerate}[label=(\alph*)]
            \item $t_\mathrm{max} = 32000$.
            \item $t_\mathrm{max} = 4000$.
        \end{enumerate}
    \item \textbf{Figure 3:} $t_\mathrm{max} = 4000$.
    \item \textbf{Figure 4:} All panels $N_c=10^{-15}$.
    \begin{enumerate}[label=(\alph*)]
    \item Top: $\mu=2, g=1, D=8, T=2$. Bottom: $\mu=4, g=2, D=2, T=6, P=2^{24}$.
    \item $S=512, P=2^{16}$, $t_\mathrm{max} = 8000$.
    \item $\mu=2, g=1, \theta=0.3, K=1, t_\mathrm{max} = 1000$.
    \item Same as (c) but $g=0.5$.
    \end{enumerate}
\end{itemize}

\bibliography{bibliography}